\documentclass[aps,amsmath,amssymb,reprint,groupedaddress]{revtex4-1}
\usepackage{graphicx} % 
\usepackage{subfigure}   % ²åÈësubfigure ÐèÒª

\begin{document}

%\preprint{}

\title{Electro-Neutral Models for Dynamic Poisson-Nernst-Planck System}


\author{Zilong Song, Xiulei Cao, Huaxiong Huang}
%\thanks{}
\affiliation{Department of Mathematics and Statistics, York University and Fields Institute for Research in Mathematical Sciences, Toronto, Ontario, Canada}


\date{\today}

\begin{abstract}
The Poisson-Nernst-Planck (PNP) system is a standard model for describing ion transport. In many applications, e.g., ions in biological tissues, the presence of thin boundary layers poses both modelling and computational challenges. In this paper,    
we derive simplified electro-neutral (EN) models where the thin boundary layers are replaced by effective boundary conditions. There are two major advantages of EN models. First of all, it is much cheaper to solve them numerically. Secondly, EN models are easier to deal with compared with the original PNP, therefore it is also easier to derive macroscopic models for cellular structures using EN models.  

Even though the approach is applicable to higher dimensional cases, this paper mainly focuses on the one-dimensional system, including the general multi-ion case. Using systematic asymptotic analysis, we derive a variety of effective boundary conditions directly applicable to EN system for the bulk region. This EN system can be solved directly and efficiently without computing the solution in the boundary layer. The derivation is based on matched asymptotics, and the key idea is to bring back higher order contributions into effective boundary conditions. For Dirichlet boundary conditions, the higher order terms can be neglected and classical results (continuity of electrochemical potential) are recovered. For flux boundary conditions, however, neglecting higher contribution leads to physically incorrect solutions since they account for accumulation of ions in boundary layer. The validity of our EN model is verified by several examples and numerical computation. In particular, our EN model is much more efficient than the original PNP model when applied to the computation of membrane potential. Implemented with the Hodgkin-Huxley model, the computational time for solving the EN model is significantly reduced without sacrificing the accuracy of the solution due to the fact that it allows for relatively large mesh and time step sizes.
\end{abstract}

%\keywords{}

\maketitle

\section{Introduction}

The Poisson-Nernst-Planck (PNP) system describes the transport of ions under the influence of both an ionic concentration gradient and an electric field.  It is essentially a system coupling diffusion and electrostatics, and the nonlinearity comes from the drift effect of electric field on ions. Such a system and its variants have found extensive and successful applications in biological systems, in particular in the description of ion transport through cells and ion channels \cite{bob2001,chun2012}. It has also been applied to many industrial fields, such as the semiconductor devices \cite{markowich2013} and the detection of poisonous lead by ion-selective electrode \cite{lead2013}.

One intriguing feature of this system is the presence of boundary layer (BL) near the boundary of concerned domain, often called Debye layer in literature. A large number of works have been devoted to the BL analysis of PNP systems. For example, singular perturbation analysis of PNP system has been carried out for narrow ion channels with certain geometric structure \cite{bob2008,singer2009}. Geometric singular perturbation approach has been developed to investigate the existence and uniqueness of solutions in stationary PNP system \cite{liu2009,liu2015} as well as the effects of permanent charge and ion size \cite{liu2013,liu2015b}. Recently, Wang et al. \cite{xiangsheng2014} have tackled the steady state PNP system with arbitrary number of ion species and arbitrary valences, and have successfully reduced the asymptotic solutions to a single scalar transcendental equation.

In general, the solution consists of two parts, the BL solution in a small neighbourhood of  boundary and the bulk solution in the interior region of the domain. In one-dimensional (1D) case, the leading order solution in BL can be constructed either explicitly or in integral form. Based on the BL analysis, effective continuity conditions have been proposed to connect the bulk solution and BL solution, e.g., the continuity of electro-chemical potential in \cite{rubinstein1990}. These effective conditions have been applied to the study of steady states of 1D systems, showing the existence of multiple steady states with piecewise constant fixed charge \cite{rubinstein1987}. One objective of this paper is to generalize the effective conditions for other boundary conditions. These conditions replace the BL region and have potential applications for deriving macroscopic models \cite{huang2011} of bulk region in complicated structures. For example, some macroscopic continuum equations are derived in bulk region for the lens circulation \cite{lens1985,lens2013}, by taking into account the fluxes through membranes with an ad hoc model for the BL effect, so the fluxes calculated there might not be accurate.

The other objective of this paper is related to numerical computation of PNP systems. In addition to the BL analysis, many (conservative) numerical schemes have been developed for PNP systems, such as finite element method \cite{gao2017}, finite difference scheme \cite{liu2014} and finite volume method \cite{chainais2003,chainais2003b}, in one or high-dimensional spaces~\cite{lu2010,mirzadeh2014}. Due to the presence of BL, computation of the PNP system needs to accurately capture the behaviour of solution in BL. Since the solutions change rapidly in BL, more mesh points are needed in BL than in the bulk region to attain certain accuracy, requiring advanced techniques such as adaptive refined mesh and moving mesh \cite{weizhanghuang1996,tang2003}. In general, computational cost is higher and development of numerical method is more demanding, especially when there are many BLs in a complicated structure. Having effective formulas/conditions to replace the BL will significantly reduce the computational time as well as the effort for developing sophisticated numerical methods, since under such framework the solutions in the bulk region can be obtained directly. 

In this work, we will focus on the 1D dynamic PNP system and derive an electro-neutral (EN) system for bulk region with effective boundary conditions for several boundary conditions. In Section II, we first present the formulation for the two-ion species case and related EN models, with Dirichlet or flux boundary conditions for ion concentration and Dirichlet or Robin boundary conditions for electric potential. A more general multi-ion species model is presented afterwards. In Section III, these effective boundary conditions are validated by one steady state and two dynamic examples. In Section IV, we combine the PNP system with the Hodgkin-Huxley model and derive an EN model for neuronal axon, capturing the phenomenon of action potential efficiently. Finally conclusions and discussion of future directions are given in Section V.

\section{The electro-neutral theories}

In this section, we present the electro-neutral (EN) systems with various effective boundary conditions. To introduce the main ideas, we first present the simplest PNP system, for $\pm 1$ ion specices, where the solutions and effective boundary conditions are explicit. It is followed by the general multi-ion species case.

We consider the 1D dynamic PNP system in the normalized interval $0<x<1$,
\begin{equation}
\label{eq1}
\begin{aligned}
&- \epsilon^2 \partial_{xx}\psi  = p-n,\\
&\partial_t p = - \partial_{x} (J_p)=\partial_{x} (\partial_{x}p + p \partial_{x}\psi),\\
&\partial_t n = - \partial_{x}(J_n)=\partial_{x}(\partial_{x}n - n \partial_{x}\psi),
\end{aligned}
\end{equation}
where $p(x,t)$ and $n(x,t)$ are the concentrations of the two ions with valences $\pm1$, $J_p$ and $J_n$ are the associated two fluxes of positive and negative ions, $\psi(x,t)$ denotes the electric potential and $\epsilon \ll 1$ is a small parameter (a combination of dielectric constant and other constants). The diffusion constants have been assumed to be 1 for simplicity, since it does not cause essential difference. Generalization to the multi-ion case with different diffusion constants will be mentioned later. We will consider various types of boundary conditions at two ends in the following subsections. For example, as in Subsection B, we can adopt Dirichlet condition for $\psi$ and two flux conditions at $x=0$
\begin{equation}
\label{eq2}
\begin{aligned}
\psi(0,t) = \psi_0(t), \,  J_p(0,t) = J_{p,0}(t), \, J_n(0,t) = J_{n,0}(t).
\end{aligned}
\end{equation}
We will also replace flux conditions by Dirichlet conditions of concentration (in Subsection C) and Dirichlet condition of $\psi$ by Robin-type condition (in Subsection D). The treatment will be similar at the other end $x=1$. To complete the system, we also need the initial conditions $p(x,0), n(x,0)$ for two ions. But the initial effect is not considered in this work, and we mainly limit ourselves to the large time behaviour of solutions (when BL is already present) or the case near equilibrium state.

We focus on the case when local electro-neutrality (LEN) condition in bulk region is satisfied, and there is no extra $O(1)$ unbalanced charge present in the system/interval, or more precisely there is only $O(\epsilon)$ unbalanced charge, here called near global electro-neutrality (NGEN) condition. We will illustrate later what kinds of boundary conditions fall in this case. These conditions can be justified in many biological applications, for example in the neuronal axon \cite{hodgkin1990}. Thus, in the bulk region, we assume all the functions concerned and their derivatives are $O(1)$, i.e.,
\begin{equation}
\label{eq3}
\begin{aligned}
&\psi,  \partial_{x}\psi, \partial_{xx}\psi \sim O(1), \quad p, \partial_{t}p,\partial_{x}p,... \sim O(1),\\
& n, \partial_{t} n,\partial_{x} n,... \sim O(1).
\end{aligned}
\end{equation}
Then, we obtain approximately the electro-neutral condition $p \approx n$ from the first equations in (\ref{eq1}) and more precisely we write
\begin{equation}
\label{eq4}
\begin{aligned}
& p(x,t;\epsilon) = c(x,t;\epsilon) +O(\epsilon^2),\\
& n(x,t;\epsilon) = c(x,t;\epsilon) + O(\epsilon^2),\\
&\psi(x,t;\epsilon) = \phi(x,t;\epsilon) +O (\epsilon^2),
\end{aligned}
\end{equation}
where $c$ and $\phi$ may depend on $\epsilon$ due to boundary conditions, in other words $c$ and $\phi$ can contain $O(\epsilon)$ terms if boundary conditions have such terms.
Substituting into second and third equation in (\ref{eq1}) gives the EN equations
\begin{equation}
\label{eq5}
\begin{aligned}
&\partial_t c = \partial_{x}\left( \partial_{x}c + c \partial_{x}\phi \right) + O(\epsilon^2),\\
&\partial_t c = \partial_{x}\left(\partial_{x} c - c \partial_{x} \phi \right) +O(\epsilon^2).
\end{aligned}
\end{equation}
By addition and subtraction, we can also write them as
\begin{equation}
\label{eq6}
\begin{aligned}
\partial_t c = \partial_{xx}c + O(\epsilon^2),\quad \partial_{x}(c\, \partial_{x}\phi) =0+ O(\epsilon^2).
\end{aligned}
\end{equation}
To complete this system, two effective boundary conditions are needed instead of the original three. Based on the behavior of BL solutions, we aim to derive effective conditions that connect real boundary values of $p,n,\psi$ (or boundary fluxes) and limit boundary values of bulk solution $c,\phi$ (or bulk fluxes). Finally we get a EN system for $c,\phi$ in the bulk region, which can be solved directly. In the following, we will always take $x=0$ for example, and briefly state the results for the other end.

\subsection{The leading order solution in BL}
\label{sec2_1}

From some steady state analysis, e.g. with finite fluxes or Dirichlet conditions in \cite{rubinstein1990} and  for Poisson-Boltzmann type equations in \cite{Lee2010} in the absence of extra $O(1)$ unbalanced charge, there is boundary layer with thickness $O(\epsilon)$. Also, some numerical evidence shows that, for finite fluxes, as long as the NGEN condition is satisfied, the system has BLs near end points with all $p,n,\psi$ being $O(1)$. In this subsection, we present the leading order solutions for the PNP system.  Although the solutions are well-known in literature, we give a brief derivation to be self-contained and to be more clear about the remainders. 

In the BL near $x=0$, we assume 
\begin{equation}
\label{eq7}
\begin{aligned}
& \psi,n,p \sim O(1), \quad \partial_t p, \partial_t n \sim O(1),\quad J_n, J_p \sim O(1),\\ 
&\partial_x \psi, \partial_x p, \partial_x n\sim O({1}/{\epsilon}), \quad \partial_{xx} \psi \sim O({1}/{\epsilon^2}),\end{aligned}
\end{equation}
and thus we set
\begin{equation}
\label{eq8}
\begin{aligned}
\Phi (X) = \psi(x), \, N(X)= n(x), \, P(X) = p(x),\, X= \frac{x}{\epsilon},
\end{aligned}
\end{equation}
where the argument $t$ is omitted in above functions. Then the system of equations in BL is 
\begin{equation}
\label{eq9}
\begin{aligned}
&- \partial_{XX}\Phi  = P- N, \\
& \partial_{X} ( \partial_{X}P + P \partial_{X} \Phi) = \epsilon^2 \partial_t P = O(\epsilon^2),\\
& \partial_{X} ( \partial_{X}N - N \partial_{X} \Phi) = \epsilon^2 \partial_t N = O(\epsilon^2).\\
\end{aligned}
\end{equation}
Integrating the second and third equations once, we get
\begin{equation}
\label{eq10}
\begin{aligned}
& \partial_{X} P + P \partial_{X}\Phi= - \epsilon J_{p,0} + O(\epsilon^2),\\
& \partial_{X} N - N \partial_{X} \Phi= - \epsilon J_{n,0} + O(\epsilon^2),\\
\end{aligned}
\end{equation}
where $J_{p,0},J_{n,0}$ are the finite fluxes at $x=0$.
We denote $c_0,\phi_0$ as the limit values of bulk solutions $c(x),\phi(x)$ at $x=0$, and they should match $P(\infty),\Phi(\infty)$ to leading order, implying
\begin{equation}
\label{eq11}
\begin{aligned}
& P(X) = c_0 e^{ \phi_0 - \Phi(X)} + O(\epsilon), \\
&  N(X) = c_0   e^{ \Phi(X) - \phi_0 } + O(\epsilon).
\end{aligned}
\end{equation}
Substituting into the first equation of (\ref{eq9}), we get the Poisson-Boltzmann equation as leading order equation for $\Phi$
\begin{equation}
\label{eq12}
\begin{aligned}
- \partial_{XX} \Phi  =  c_0  \left( e^{ \phi_0 - \Phi(X)} -  e^{ \Phi(X) - \phi_0  }\right) + O(\epsilon).
\end{aligned}
\end{equation}
This can be integrated out by using $\partial_X \Phi (\infty) \rightarrow 0$ and $\Phi (0) = \psi_0(t)$ (suppose it is known here or can be expressed by known boundary conditions). Finally, we obtain
\begin{equation}
\label{eq13}
\begin{aligned}
 \Phi (X) = \phi_0 + 2 \ln \frac{1- e^{-\sqrt{2 c_0 } X } \tanh\left(\frac{\phi_0  - \psi_0}{4}\right)}{1+ e^{-\sqrt{2 c_0 } X } \tanh\left(\frac{\phi_0  - \psi_0}{4}\right)} +O(\epsilon).
\end{aligned}
\end{equation}
And the solutions for $P(X,t),N(X,t)$ become
\begin{equation}
\label{eq14}
\begin{aligned}
P(X) = c_0 \left( \frac{1+  e^{-\sqrt{2c_0}X} \tanh\left(\frac{\phi_0 - \psi_0}{4}\right)}{1- e^{-\sqrt{2c_0}X} \tanh\left(\frac{\phi_0 - \psi_0}{4}\right)}\right)^2 + O(\epsilon),\\
N(X) = c_0 \left( \frac{1-  e^{-\sqrt{2c_0 }X} \tanh\left(\frac{\phi_0 - \psi_0}{4}\right)}{1+ e^{-\sqrt{2c_0}X} \tanh\left(\frac{\phi_0 - \psi_0}{4}\right)}\right)^2 + O(\epsilon).
\end{aligned}
\end{equation}
Note that in general $c_0,\phi_0,\psi_0$ are functions of $t$. The composite solutions are given by 
\begin{equation}
\label{eq15}
\begin{aligned}
&p(x) = P(X) + c(x)- c_0 +O(\epsilon),\\
&n(x) = N(X) + c(x)- c_0 +O(\epsilon),\\
&\psi(x) = \Phi(X) + \phi(x) - \phi_0 + O(\epsilon),
\end{aligned}
\end{equation}
which are uniformly valid for some finite interval $[0,\delta]$, say $\delta=1/2$, with remainder $O(\epsilon)$. One can also add the contribution of BL solution near $x=1$ (with transform $X=(1-x)/\epsilon$ and quantities $c_0,\phi_0,\psi_0$ being replaced by $c_1,\phi_1,\psi_1$)  to make the composite solution valid for the whole interval $[0,1]$. Since in the bulk we have $p(x) =c(x) +O(\epsilon^2)$ by (\ref{eq4}), it is reasonable to expect $p = c(x) +o(\epsilon)$ in some intermediate region $x\sim O(\epsilon^\alpha)$ with $0<\alpha<1$, particularly we may choose $\alpha=1/2$.

\subsection{Flux boundary condition}
\label{sec2_2}

In this subsection, we consider the case with the flux boundary conditions for two concentrations and Dirichlet condition for electric potential. More precisely, at $x=0$, we have
\begin{equation}
\label{eq16}
\begin{aligned}
\psi(0,t) = \psi_0(t), \, J_p(0,t) = J_{p,0}(t), \, J_n(0,t) = J_{n,0}(t),
\end{aligned}
\end{equation}
where $ \psi_0,J_{p,0}, J_{n,0}$ are given.  The objective is to propose two effective boundary conditions for $c,\phi$ at $x=0$ based on these three functions. 

To this end, we define for the EN system two fluxes
\begin{equation}
\label{eq17}
\begin{aligned}
J_{c}^{\pm}(x,t) = - (\partial_{x} c\pm c \, \partial_{x} \phi),
\end{aligned}
\end{equation}
and the limit values at $x=0$ are denoted by $J_{c,0}^{\pm}(t)$ respectively.
Based on assumptions (\ref{eq3},\ref{eq4}) in the bulk region, the two fluxes are almost the same as the two fluxes of original PNP system 
\begin{equation}
\label{eq18}
\begin{aligned}
J_{c}^+(\delta,t) = J_p (\delta,t) +O(\epsilon^2),  \, J_{c}^-(\delta,t) = J_n (\delta,t) +O(\epsilon^2),
\end{aligned}
\end{equation}
where $\delta$ is some generic point in bulk region, say $\delta = 1/2$. Next, we intend to find the connection between $J_{c,0}^+$ and $J_{p,0}$, or similarly between $J_{c,0}^-$ and $J_{n,0}$ at boundary. For this purpose, by integration of transport equations $(\ref{eq1})_2$ and $(\ref{eq5})_1$, we immediately get
\begin{equation}
\label{eq19}
\begin{aligned}
J_p(\delta,t) =& J_{p,0}(t) - \partial_t \int_0^\delta p(x,t) dx,\\
J_{c}^+(\delta,t) = &  J_{c,0}^+(t) - \partial_t \int_0^\delta c(x,t) dx.
\end{aligned}
\end{equation} 
Combining these two and utilizing the composite solution (\ref{eq15}), we obtain
\begin{equation}
\label{eq20}
\begin{aligned}
J_{c,0}^+ =& J_{p,0}  - \partial_t \int_0^\delta  (p-c) dx + O(\epsilon^2)\\ 
=& J_{p,0}  - \partial_t \left( \int_0^{\sqrt{\epsilon}}  (p-c) dx + \int_{\sqrt{\epsilon}}^\delta  (p-c) dx \right) + O(\epsilon^2)\\
=& J_{p,0}  - \partial_t \int_0^{\sqrt{\epsilon}}  (P(x/\epsilon)-c_0 ) dx + o(\epsilon)\\
=&  J_{p,0}  - \epsilon\, \partial_t \int_0^\infty  (P(X)-c_0  ) dX + o(\epsilon)\\
=&  J_{p,0}  - \epsilon \partial_t  \left( \sqrt{2 c_0 }  (e^{(\phi_0   - \psi_0)/2} -1)\right) + o(\epsilon),
\end{aligned}
\end{equation}
where we have used the assumption that $p=c+o(\epsilon)$ for $x\ge \sqrt{\epsilon}$, and by setting upper limit of integral as $\infty$ only exponentially small terms are neglected. In the above, $\phi_0,c_0,\psi_0$ may depend on $t$. Similarly for the other flux, we obtain the relation
\begin{equation}
\label{eq21}
\begin{aligned}
J_{c,0}^- =  J_{n,0}  - \epsilon \partial_t  \left(\sqrt{2 c_0 }  (e^{(\psi_0- \phi_0 )/2} -1)\right)+ o(\epsilon).
\end{aligned}
\end{equation}
Physically, the quantity $\psi_0- \phi_0$ in above formulas is often referred to as the zeta potential \cite{kirby2010}. To see clearly the two conditions, we carry out a linearization regarding small $\psi_0- \phi_0 $. In this case, they reduce to  
\begin{equation}
\label{eq22}
\begin{aligned}
 & J_{c,0}^+ - J_{p,0}  \approx \epsilon \partial_t  \left( \sqrt{ c_0 /2}  (\psi_0 - \phi_0 ) \right),\\
 & J_{c,0}^- -J_{n,0} \approx \epsilon \partial_t  \left( \sqrt{ c_0 /2}  ( \phi_0  - \psi_0) \right).\\
\end{aligned}
\end{equation}
Thus, by comparing these conditions, the total flux will almost have no difference while electric current changes, i.e.,
\begin{equation}
\label{eq23}
\begin{aligned}
& (J_{c,0}^+ + J_{c,0}^-) \approx  (J_{p,0} + J_{n,0}), \\
& (J_{c,0}^+ - J_{c,0}^-) -  (J_{p,0} - J_{n,0}) \approx 2 \epsilon \partial_t  \left( \sqrt{ c_0 /2}  (\psi_0 - \phi_0 ) \right).
 \end{aligned}
\end{equation}
Physically, this means some cations/anions accumulate in the BL, and the second formula is similar to that of a capacitor. The treatment for the other end $x=1$ is similar, and we summarise the results below.

\noindent \textit{\textbf{Proposition 1.} Suppose the LEN and NGEN conditions are satisfied, and let $\psi_0(t), J_{p,0}(t),  J_{n,0}(t)$ be the given electric potential and ion fluxes at $x=0$ as in (\ref{eq16}) and let $\psi_1(t), J_{p,1}(t),  J_{n,1}(t)$ be given at $x=1$ for original system (\ref{eq1}), then we have the effective boundary conditions for the EN system (\ref{eq6}) 
\begin{equation}
\label{eq24}
\begin{aligned}
& J_{c,0}^+ =  J_{p,0}  - \epsilon \partial_t  \left( \sqrt{2 c_0 }  (e^{(\phi_0   - \psi_0)/2} -1)\right) + o(\epsilon),\\
& J_{c,0}^- =  J_{n,0}  - \epsilon \partial_t  \left(\sqrt{2 c_0 }  (e^{(\psi_0- \phi_0 )/2} -1)\right)+ o(\epsilon),\\
& J_{c,1}^+ =  J_{p,1}  +  \epsilon \partial_t  \left(\sqrt{2 c_1}  (e^{(\phi_1 -\psi_1)/2} -1)\right)+ o(\epsilon),\\
& J_{c,1}^- =  J_{n,1}  + \epsilon \partial_t  \left(\sqrt{2 c_1 }  (e^{(\psi_1- \phi_1 )/2} -1)\right)+ o(\epsilon),
 \end{aligned}
\end{equation}
where $J_c^{\pm}$ are defined by (\ref{eq17}) and subscripts 0 and 1 denote quantities at $x=0$ and $x=1$ respectively.
}

\noindent {\bf Remark 1.} Keeping the $O(\epsilon)$ terms in the formula (\ref{eq24}) is necessary for two reasons: first,  in bulk equations (\ref{eq5}) we have assumed an $O(\epsilon^2)$ remainder so it is reasonable and consistent to bring back the $O(\epsilon)$ terms on boundary conditions; second, neglecting the $O(\epsilon)$ terms is physically incorrect for EN system as the solution would not be unique (e.g., $\phi$ can differ by a constant). \\
\noindent {\bf Remark 2.} In this case, the fluxes $J_{p,0},J_{n,0}$ can be either $O(1)$ or $O(\epsilon)$, as long as the NGEN is satisfied.  This means when fluxes are $O(1)$, we should require the fluxes are almost balanced $J_{p,1} - J_{p,0}= J_{n,1} - J_{n,0}+O(\epsilon)$ or its integral over time satisfies
\begin{equation}
\label{eq24_1}
\begin{aligned}
\int_0^t (J_{p,1} - J_{p,0}) dt= \int_0^t (J_{n,1} - J_{n,0}) dt+ O(\epsilon).
 \end{aligned}
\end{equation}
Otherwise, the solution in BL will not be $O(1)$ anymore. For a steady state (Poisson-Boltzmann type equation  in \cite{Lee2010}) with extra $O(1)$ unbalanced charge, the solution $\psi$ in BL is shown to have a span of $O(\log(1/\epsilon))$.

\subsection{Dirichlet boundary condition revisited}

In this subsection, we will consider the case with Dirichlet boundary conditions for two ion concentrations. We also take left end $x=0$ for example, and have 
\begin{equation}
\label{eq25}
\begin{aligned}
\psi(0,t) = \psi_0(t), \, p(0,t) = p_0(t), \, n(0,t)=n_0(t).
\end{aligned}
\end{equation}
The leading order effective boundary conditions for this case are well-known. With the same assumptions as previous subsection, we arrive at the same BL system, and easily get
\begin{equation}
\label{eq26}
\begin{aligned}
\partial_{X} P + P \partial_{X}\Phi=O(\epsilon),\quad \partial_{X} N - N \partial_{X} \Phi=  O(\epsilon).
\end{aligned}
\end{equation}
By integration and using the matching condition, we obtain
\begin{equation}
\label{eq27}
\begin{aligned}
& \ln c_0  + \phi_0   = \ln p_0 + \psi_0 +  O(\epsilon),\\
& \ln c_0  -  \phi_0 = \ln n_0 - \psi_0  +  O(\epsilon).
\end{aligned}
\end{equation}
These connection conditions are referred to as \textit{continuity of electro-chemical potential}, widely adopted in literature \cite{rubinstein1990}. And equivalently, the explicit effective boundary conditions for EN system are
\begin{equation}
\label{eq28}
\begin{aligned}
c_0  = \sqrt{p_0 n_0}, \quad \phi_0  = \psi_0 + \frac{1}{2} \ln (p_0/n_0).
\end{aligned}
\end{equation}

As in the bulk assumption, we would like to keep the $O(\epsilon)$ effect/terms, thus a natural question is how to bring back such $O(\epsilon)$ effect in the effective boundary conditions for the reduced EN system. One may want to seek a general expansion to $O(\epsilon)$ in BL and assume
\begin{equation}
\label{eq29}
\begin{aligned}
&\Phi= \Phi_0 + \epsilon \Phi_1 +..., \quad P = P_0 + \epsilon P_1+..., \\
& N = N_0 + \epsilon N_1+...
\end{aligned}
\end{equation}
The leading order solutions $\Phi_0,P_0,N_0$ can be immediately written down, which are the same as those in (\ref{eq13},\ref{eq14}) with replacement given by (\ref{eq28}). However, getting the explicit expression for $\Phi_1,P_1,N_1$ seems difficult. Therefore, instead we try to avoid such a solving process and intend to find the higher-order contributions directly based on leading order solution. 

Now, we take $P(X)$ for illustration, where the argument $t$ is omitted here and in the following. The second equation in BL system implies
\begin{equation}
\label{eq30}
\begin{aligned}
 \partial_X P + P  \partial_X \Phi = - \epsilon J_{p,0} +O(\epsilon^2 X),
\end{aligned}
\end{equation}
where $J_{p,0}$ is some unknown flux constant. Dividing both sides by $P$, we get
\begin{equation}
\label{eq31}
\begin{aligned}
 \partial_X (\ln P +  \Phi)=& - \epsilon J_{p,0} /P +O(\epsilon^2 X)\\
 =& - \epsilon J_{p,0} /P_0 +O(\epsilon^2 X).
\end{aligned}
\end{equation}
From previous subsection, we know that $J_{p,0} = J_{c,0}^+ + O(\epsilon)$. Therefore, we obtain
\begin{equation}
\label{eq32}
\begin{aligned}
 &\ln (P(X)) +  \Phi (X)  \\
 =&   \ln p_0 +  \psi_0  - \epsilon J_{c,0}^+  \int_0^X 1/P_0(z) dz +O(\epsilon^2 X).
\end{aligned}
\end{equation}
By matching \cite{bush1992}, let $X= \epsilon^{\alpha-1} s$ or $x=\epsilon^\alpha s$ with $1/2<\alpha<1$, we can expect that
\begin{equation}
\label{eq33}
\begin{aligned}
& P(\epsilon^{\alpha-1} s) = c (\epsilon^\alpha s) +o(\epsilon),\\
& \Phi(\epsilon^{\alpha-1} s) = \phi (\epsilon^\alpha s) +o(\epsilon).
\end{aligned}
\end{equation}
Substituting $X= \epsilon^{\alpha-1} s$ into previous relation (\ref{eq32}), we get from the left-hand side
\begin{equation}
\label{eq34}
\begin{aligned}
 & \ln (P(X)) +  \Phi (X) \\
 = &  \ln (c_0 ) + \phi_0   + \left(\frac{c'(0)}{c_0 } + \phi'(0)\right) \epsilon^\alpha s +  o(\epsilon)
\end{aligned}
\end{equation}
and from the integral on the right-hand side
\begin{equation}
\label{eq35}
\begin{aligned}
\epsilon \int_0^X \frac{1}{P_0(z)} dz =\frac{\epsilon^\alpha s}{c_0 }  + \frac{\sqrt{2} \epsilon}{c_0^{3/2} }  \left(e^{\frac{\psi_0 - \phi_0 }{2}} -1\right)  + o(\epsilon).
\end{aligned}
\end{equation}
Since $J_{c,0}^+ = -(\partial_x c(0)+ c_0 \partial_x \phi(0))$ by definition (\ref{eq17}), the $\epsilon^\alpha s$ terms automatically cancel each other (which partially verifies the correctness of matching), and we are left with
\begin{equation}
\label{eq36}
\begin{aligned}
& \ln c_0  + \phi_0   +  \frac{\sqrt{2}  J_{c,0}^+  \epsilon}{c_0^{3/2} } \left(e^{({\psi_0 - \phi_0 })/{2}} -1\right) \\
&= \ln p_0 + \psi_0 +  o(\epsilon) .
\end{aligned}
\end{equation}
Compared with previous leading order condition (\ref{eq27}), there is an $O(\epsilon )$ correction term in above formula, so it can be considered as a generalization of continuity of electro-chemical potential. Treatments for the other condition and two conditions at $x=1$ are similar, and we summarize the results as follows.

\noindent \textit{\textbf{Proposition 2.}. Suppose the LEN and NGEN conditions are satisfied, and let $\psi_0(t), p_0(t),  n_0(t)$ be the given electric potential and ion concentrations at $x=0$ as in (\ref{eq25}) and let $\psi_1(t), p_1(t),  n_1(t)$ be given at $x=1$ for original system (\ref{eq1}), then we have the effective boundary conditions for the EN system (\ref{eq6}) 
\begin{equation}
\label{eq38}
\begin{aligned}
& \ln c_0  + \phi_0   +  \frac{\sqrt{2}  J_{c,0}^+  \epsilon}{c_0^{3/2} } \left(e^{({\psi_0 - \phi_0 })/{2}} -1\right) = \ln p_0 + \psi_0 +  o(\epsilon),\\
& \ln c_0  -  \phi_0  +  \frac{\sqrt{2}  J_{c,0}^-  \epsilon}{c_0^{3/2} } \left(e^{({\phi_0- \psi_0 })/{2}} -1\right) = \ln n_0 - \psi_0  +  o(\epsilon),\\
&\ln c_1  + \phi_1   -  \frac{\sqrt{2}  J_{c,1}^+  \epsilon}{c_1^{3/2} } \left(e^{({\psi_1 - \phi_1 })/{2}} -1\right) = \ln p_1 + \psi_1 +  o(\epsilon),\\
& \ln c_1  -  \phi_1  -  \frac{\sqrt{2}  J_{c,1}^-  \epsilon}{c_1^{3/2} } \left(e^{({\phi_1- \psi_1})/{2}} -1\right) = \ln n_1 - \psi_1  +  o(\epsilon),
\end{aligned}
\end{equation}
where $J_c^{\pm}$ are defined by (\ref{eq17}) and subscripts 0 and 1 denote quantities at $x=0$ and $x=1$ respectively.
}

\noindent \textbf{Remark 3.}  We can alternatively derive an asymptotically equivalent expressions
\begin{equation}
\label{eq40}
\begin{aligned}
c_0  = & \sqrt{p_0 n_0} + \frac{\epsilon}{\sqrt{2}} \Big\{  \left( n_0^{-1/4} - p_0^{-1/4}\right)^2 \partial_x c(0,t) \\
& + \left( \sqrt{n_0} - \sqrt{p_0} \right) \partial_x \phi(0,t) \Big\}, \\
\phi_0  = & \psi_0 + \frac{1}{2} \ln (p_0/n_0) +  \frac{\epsilon}{\sqrt{2}} \Big\{ \frac{\sqrt{n_0} - \sqrt{p_0}}{n_0 p_0} \partial_x c(0,t) \\
& +  \left( n_0^{-1/4} - p_0^{-1/4}\right)^2 \partial_x \phi(0,t) \Big\},\\
c_1  =&  \sqrt{p_1 n_1} - \frac{\epsilon}{\sqrt{2}} \Big\{  \left( n_1^{-1/4} - p_1^{-1/4}\right)^2 \partial_x c(1,t) \\
& + \left( \sqrt{n_1} - \sqrt{p_1} \right) \partial_x \phi(1,t) \Big\}, \\
\phi_1  = & \psi_1 + \frac{1}{2} \ln (p_1/n_1) -  \frac{\epsilon}{\sqrt{2}} \Big\{ \frac{\sqrt{n_1} - \sqrt{p_1}}{n_1 p_1} \partial_x c(1,t) \\
& +  \left( n_1^{-1/4} - p_1^{-1/4}\right)^2 \partial_x \phi(1,t) \Big\},
\end{aligned}
\end{equation}
which are Robin-type boundary conditions. Compared with (\ref{eq28}), there are $O(\epsilon)$ corrections in above conditions. For a special case, say $n_0= p_0 +O(\epsilon)$ at $x=0$, the correction terms will be of higher order, then continuity of electro-chemical potential (\ref{eq28}) holds with remainder $o(\epsilon)$. 

\noindent \textbf{Remark 4.} In some cases, $p(0,t)= p_0(t)$ is not explicitly given, but is related to flux $J_{p,0}$, so proper modification is needed. For example in biological applications, there is certain relation between flux and ion concentration across cell membrane or ion channel, such as Hodgkin-Huxley model \cite{hodgkin1990} or GHK flux model \cite{hille2001}. And, in electrolyte there is the Chang-Jaffle boundary condition \cite{chang1952,chang2015} or modified Chang-Jaffle condition \cite{lead2013}. Suppose the boundary condition is in the form $J_{p,0} = f(p_0)$, where $f$ is some given function, then we need to supplement the two conditions at $x=0$ with 
\begin{equation}
\label{eq39}
\begin{aligned}
 f(p_0) = J_{c,0}^+ + \epsilon \partial_t  \left( \sqrt{2 c_0 }  (e^{(\phi_0   - \psi_0)/2} -1)\right).
\end{aligned}
\end{equation}
If mixed conditions are given (e.g., one Dirichlet and one flux), then we need to combine the two relevant boundary conditions from the two propositions, e.g., if $p_0$ and $J_{n,0}$ are given, we should use $(\ref{eq38})_1$ and $(\ref{eq24})_2$.

\subsection{Robin-type boundary condition for $\psi$}
\label{sec2_4}

In this subsection, we consider the case when Dirichlet condition of electric potential $\psi$ is replaced by Robin-type boundary condition. The Robin-type condition is often used to model the property of membrane or the stern layer near boundary. The previous effective conditions need to be modified since the quantity $\psi_0$ in those formulas is unknown. 

\subsubsection{Dirichlet conditions for two ion concentrations}

Suppose we have the boundary conditions at $x=0$
\begin{equation}
\label{eq41}
\begin{aligned}
& \eta \partial_x \psi(0,t) = \psi(0,t) - g_0(t), \\
& p(0,t) = p_0(t), \quad n(0,t)=n_0(t),
\end{aligned}
\end{equation}
where $\eta$ is a parameter which is assumed to be $\eta \le O(\epsilon)$, and $g_0$ is some given function.

With previous assumptions, we still have (\ref{eq11}) and (\ref{eq12}). Integrating once and using $\partial_X \Phi(\infty) = 0$, we obtain
\begin{equation}
\label{eq42}
\begin{aligned}
(\partial_X \Phi)^2  =  2 c_0  \left( e^{ \phi_0 - \Phi(X)} +  e^{ \Phi(X) - \phi_0  } -2\right) +O(\epsilon),
\end{aligned}
\end{equation}
or equivalently
\begin{equation}
\label{eq42_1}
\begin{aligned}
\partial_X \Phi =\sqrt{2 c_0 }  \left( e^{\frac{ \phi_0 - \Phi(X)}{2}} -  e^{ \frac{\Phi(X) - \phi_0}{2}} \right) + O(\epsilon).
\end{aligned}
\end{equation}
On account of the identity $\partial_X \Phi = \epsilon \partial_x \psi$, the above condition at $x=0$ becomes
\begin{equation}
\label{eq43}
\begin{aligned}
\epsilon \partial_x \psi(0,t) =& \sqrt{2 c_0 }  \left( e^{\frac{ \phi_0 - \psi_0}{2}} -  e^{ \frac{\psi_0 - \phi_0}{2}} \right) + O(\epsilon)\\
= & \sqrt{2 p_0} -\sqrt{2 n_0} +O(\epsilon).
\end{aligned}
\end{equation}
Combining with the Robin-type condition $(\ref{eq41})_1$ leads to
\begin{equation}
\label{eq44}
\begin{aligned}
\psi_0 \equiv \psi (0,t)& = \frac{\eta}{\epsilon} ( \sqrt{2 p_0} -\sqrt{2 n_0}) + g_0 + O(\eta).
\end{aligned}
\end{equation}
We conclude that we have the same effective conditions as before 
\begin{equation}
\label{eq45}
\begin{aligned}
& \ln c_0  + \phi_0   +  \frac{\sqrt{2}  J_{c,0}^+  \epsilon}{c_0^{3/2} } \left(e^{({\psi_0 - \phi_0 })/{2}} -1\right) \\
=&  \ln p_0 + \psi_0 +  o(\epsilon)+ O(\eta),\\
 & \ln c_0  -  \phi_0  +  \frac{\sqrt{2}  J_{c,0}^-  \epsilon}{c_0^{3/2} } \left(e^{({\phi_0- \psi_0 })/{2}} -1\right)\\
 =&  \ln n_0 - \psi_0  +  o(\epsilon) + O(\eta),
\end{aligned}
\end{equation}
except that $\psi_0$ is given by (\ref{eq44}) in this case.

Note that if $\eta=O(\epsilon)$, we can omit the $O(\epsilon)$ terms in above conditions since they are not exact. If $\eta = o(\epsilon)$,  the condition will become close to that in the Dirichlet case for $\psi$. In particular, for $\eta = o(\epsilon^2)$, the ${\eta}/{\epsilon}$ term can be neglected in (\ref{eq44}), which essentially reduces to the Dirichlet case $\psi (0,t)= g_0$ (see (\ref{eq41})). If $\eta/\epsilon $ tends to infinity (not considered here), the previous BL assumptions might not be true unless $n_0\approx p_0$, and this is left for future study. 

The treatment for the right end $x=1$ is similar and we summarize the results below.

\noindent \textit{\textbf{Proposition 3.} Suppose for original system (\ref{eq1}), the assumptions and conditions are the same as Proposition 2 except that the conditions for $\psi_0(t), \psi_1(t)$ are replaced by 
\begin{equation}
\label{eq46}
\begin{aligned}
& \eta \partial_x \psi(0,t) = \psi(0,t) - g_0(t), \\
& \eta \partial_x \psi(1,t) = g_1 (t) - \psi(1,t),
\end{aligned}
\end{equation}
where $\eta \le O(\epsilon)$, then we have the same effective boundary conditions for the EN system (\ref{eq6}) as Proposition 2 except that $\psi_0(t), \psi_1(t)$ in (\ref{eq38}) are calculated by}
\begin{equation}
\label{eq48}
\begin{aligned}
\psi_i \equiv \psi (i,t)& = \frac{\eta}{\epsilon} ( \sqrt{2 p_i} -\sqrt{2 n_i}) + g_i + O(\eta),
\end{aligned}
\end{equation}
where $i=0,1$.

\subsubsection{Flux conditions}

For this case, the boundary conditions at $x=0$ are of the form
\begin{equation}
\label{eq49}
\begin{aligned}
 & \eta \partial_x \psi(0,t) = \psi(0, t) - g_0(t), \\
 & J_p(0,t) = J_{p,0}(t), \quad J_n(0,t) = J_{n,0}(t),
\end{aligned}
\end{equation}
where $\eta \le O(\epsilon)$. The manipulation follows similar lines as before, and we summarize the results below. 

\noindent \textit{\textbf{Proposition 4.} Suppose for original system (\ref{eq1}), the assumptions and conditions are the same as Proposition 1 except that the conditions for $\psi_0(t), \psi_1(t)$ are replaced by 
\begin{equation}
\label{eq46}
\begin{aligned}
& \eta \partial_x \psi(0,t) = \psi(0,t) - g_0(t), \\
& \eta \partial_x \psi(1,t) = g_1 (t) - \psi(1,t),
\end{aligned}
\end{equation}
where $\eta \le O(\epsilon)$, then we have the same effective boundary conditions for the EN system (\ref{eq6}) as Proposition 1 except that $\psi_0(t), \psi_1(t)$ in (\ref{eq38}) are determined by the nonlinear algebraic equation}
\begin{equation}
\label{eq51}
\begin{aligned}
\psi_i - g_i & = \frac{\eta}{\epsilon} \sqrt{2 c_i } \left( e^{\frac{\phi_i - \psi_i }{2}} -  e^{ \frac{\psi_i - \phi_i }{2}} \right) + O(\eta),
\end{aligned}
\end{equation}
where $i=0,1$.

\noindent \textbf{Remark 5.}  For the steady case, the above algebraic equation is the same as formula (1.23) in \cite{Lee2010}, with substitution $g_0 = \phi_0(1), \phi_0 = 0, \psi_0 = \phi^\ast,c_0 = \alpha/2$. With $\eta = o(\epsilon)$, the effective boundary conditions reduce to those for Dirichlet case, with $\psi(0,t) = g_0$. \\
\noindent \textbf{Remark 6.}  The case $O(\epsilon)<\eta \le O(1)$ is not considered above, since the boundary layer structure would be different. For the NGEN case considered here, it is expected that there is still a BL with thickness $O(\epsilon)$, and the above relation (\ref{eq51}) implies that $\psi_i - \phi_i= o(1)$ in BL. We postulate that in BL near $x=0$,
\begin{equation}
\label{eq53}
\begin{aligned}
& \psi - \phi_0, p -c_0,n-c_0 =O(\epsilon/\eta),\\
& \partial_x \psi,\partial_x p, \partial_x n= O(1/\eta),\quad \partial_{xx} \psi = O(1/(\eta \epsilon)),...,
\end{aligned}
\end{equation}
and it is proper to adopt the scaling $x=X/\epsilon$ and the transform $\Phi(X) = \psi - \phi_0, P(X)= p-c_0, N(X)= n-c_0$ instead. This seems true for Poisson-Boltzmann type equations in \cite{Lee2010} with their electro-neutral case, where boundary layer with $O(\epsilon)$ thickness gradually disappears when $\eta/\epsilon$ becomes larger. Then, the above conditions (\ref{eq51},\ref{eq24}) are still valid to leading order with new remainder $O(\epsilon)$, which will be verified by numerical examples in later sections.

\subsection{The general multi-ion species case}
\label{sec2_5}
 
In this subsection we consider the general case with $n$ species of ions. The concentrations of ions are denoted by $p_i$ with valences $z_i$ ($i=1,..,n$), where the valences are not necessarily different. The original PNP system for $p_i$ ($i=1,..,n$) and $\psi$ is given by
\begin{equation}
\label{Eq53}
\begin{aligned}
& - \epsilon^2 \partial_{xx}\psi  = \sum_{k=1}^n z_k p_k,\\
& \partial_t p_i =  - \partial_{x} (J_{p_i}) = D_i \partial_{x} (\partial_{x}p_i + z_i p_i \partial_{x}\psi),
\end{aligned}
\end{equation}
where $i=1,..,n$, and $D_i$ are some dimensionless diffusion constants. The reduced EN system for bulk region is
\begin{equation}
\label{Eq54}
\begin{aligned}
&\partial_t c_i = - \partial_{x} (J_{c_i})= D_i \partial_{x} (\partial_{x}c_i + z_i c_i \partial_{x}\phi),
\end{aligned}
\end{equation}
where $i=1,..,n$. By the LEN condition $\sum z_k c_k=0$, the last concentration $c_n$ can be expressed by previous ones. Finally, the EN system for $n$ unknowns $c_1,..,c_{n-1},\phi$ can be written as 
\begin{equation}
\label{Eq55}
\begin{aligned}
&\partial_t c_i = - \partial_{x} (J_{c_i})= D_i \partial_{x} (\partial_{x}c_i + z_i c_i \partial_{x}\phi),\\
&\sum_{k=1}^n z_k D_k \partial_{x} (\partial_{x}c_k + z_k c_k \partial_{x}\phi) =0,
\end{aligned}
\end{equation}
where $i=1,..,n-1$ and whenever $c_n$ appears we should replace it by $c_n = -\frac{1}{z_n}\sum_{k=1}^{n-1} z_k c_k$.

First, at $x=0$, we consider the boundary conditions of the type 
\begin{equation}
\label{Eq56}
\begin{aligned}
\psi(0,t) = \psi_0(t), \quad J_{p_i}(0,t) = J_{p_i,0}(t), \quad i=1,..,n.
\end{aligned}
\end{equation}

\noindent \textit{\textbf{Theorem 1.} Suppose the LEN and NGEN conditions are satisfied, and let $\psi_0(t), J_{p_i,0}(t)$ be the given electric potential and ion fluxes at $x=0$ as in (\ref{Eq56}) and let $\psi_1(t), J_{p_i,1}(t)$ be given at $x=1$ for original system (\ref{Eq53}), then for the EN system (\ref{Eq55}) we have the effective boundary conditions  
\begin{equation}
\label{Eq57}
\begin{aligned}
& J_{c_i,0} = J_{p_i,0} - \epsilon \partial_t F_i(c_{k0},\phi_0-\psi_0) + o(\epsilon),\\
& J_{c_i,1} = J_{p_i,1} + \epsilon \partial_t F_i(c_{k1},\phi_1-\psi_1) + o(\epsilon),
\end{aligned}
\end{equation}
where $i=1,..,n$, $J_{c_i} = -D_i (\partial_{x}c_i + z_i c_i \partial_{x}\phi)$ are defined in (\ref{Eq54}), the argument $c_{k0}$ represents a vector $(c_{10},..,c_{n-1,0})$, subscripts 0 and 1 denote quantities at $x=0$ and $x=1$ respectively, and 
\begin{equation}
\label{Eq58}
\begin{aligned}
& F_i(c_{k0},\phi_0-\psi_0) \\
&= \pm \frac{c_{i0}}{\sqrt{2}} \int_1^{e^{\phi_0 -\psi_0}} \frac{u^{z_i} -1}{\sqrt{\sum_{k=1}^n c_{k0} (u^{z_k} -1)}} \frac{du}{u}.
\end{aligned}
\end{equation}
Here, the $\pm$ are chosen for the cases $\psi_0\le \phi_0$ and $\psi_0\ge \phi_0$ respectively, but $F_i$ is well-defined around $\phi_0= \psi_0$, and if $F_i$ can be integrated out, the expressions from the two cases are the same. 
}

\noindent \textit{Proof.}  The derivation follows similar lines as in subsections \ref{sec2_1} and \ref{sec2_2}, and here we only mention the key steps different from previous case. Near $x=0$, with the scalings 
\begin{equation}
\label{Eq59}
\begin{aligned}
\Phi (X) = \psi(x), \quad P_i(X) = p_i(x),\quad X= \frac{x}{\epsilon},
\end{aligned}
\end{equation}
where $i=1,..,n$, and by the BL analysis, we get
\begin{equation}
\label{Eq60}
\begin{aligned}
 -\partial_{XX} \Phi = \sum_{i=1}^n z_i P_i(X)  = \sum_{i=1}^n z_i c_{i0} e^{z_i (\phi_0 - \Phi(X))}.
\end{aligned}
\end{equation}
Integrating once gives
\begin{equation}
\label{Eq61}
\begin{aligned}
\partial_{X} \Phi = \pm \sqrt{2 \sum_{i=1}^n c_{i0} \left(  e^{z_i (\phi_0 - \Phi(X))} -1 \right)},
\end{aligned}
\end{equation}
where $\pm$ are chosen for the cases $\psi_0\le \phi_0$ and $\psi_0\ge \phi_0$ respectively. Then by utilizing the transport equations, we obtain
\begin{equation}
\label{Eq62}
\begin{aligned}
&J_{c_i,0} = J_{p_i,0} - \epsilon \partial_t F_i(c_{k0},\phi_0-\psi_0) + o(\epsilon),
\end{aligned}
\end{equation}
where $F_i$ depends on all ion concentrations $c_{k0}$ ($k=0,..,n-1$, $c_{n0}$ is replaced by previous ones) and is given by
\begin{equation}
\label{Eq63}
\begin{aligned}
& F_i(c_{k0},\phi_0-\psi_0) =  \int_0^\infty (P_i(X) - c_{i0} ) dX \\
=& \pm \frac{c_{i0}}{\sqrt{2}} \int_1^{e^{\phi_0 -\psi_0}} \frac{u^{z_i} -1}{\sqrt{\sum_{k=1}^n c_{k0} (u^{z_k} -1)}} \frac{du}{u}
\end{aligned}
\end{equation}
where we have made change of variable $u= e^{\phi_0 - \Phi(X)}$, and $\pm$ are chosen for the cases $\psi_0\le \phi_0$ and $\psi_0\ge \phi_0$ respectively. 

It can be easily seen that $F_i$ is well-defined around $\phi_0= \psi_0$, in particular $F_i =0$ when $\phi_0= \psi_0$. And, if $F_i$ can be integrated out, the expressions from the two cases are the same. This is readily verified from the fact that there is a factor $(u-1)^2$ inside square root in the denominator of the integrand, which cancels with the $\pm$ sign and the factor $u-1$ in the numerator (see Appendix \ref{appendix_A} for details). \qquad $\square$

\noindent \textbf{Remark 7.} For some special cases, the explicit expressions for $F_i$ are available. For the previous case $z_1=1,z_2=-1$, we recover the result
\begin{equation}
\label{Eq64}
\begin{aligned}
& F_1(c_{10},\phi_0-\psi_0) =  \sqrt{2 c_{10}} (e^{(\phi_0 - \psi_0)/2} -1),\\
& F_2(c_{10},\phi_0-\psi_0) =  \sqrt{2 c_{10}} (e^{(\psi_0 - \phi_0)/2} -1).
\end{aligned}
\end{equation}
For the case $z_1=2,z_2 =-1$, we get
\begin{equation}
\label{Eq65}
\begin{aligned}
F_1 (c_{10},\phi_0-\psi_0)=&  \sqrt{\frac{c_{10}}{2}} \left[ e^{\frac{\phi_0 - \psi_0}{2}} \sqrt{e^{(\phi_0 - \psi_0)} +2} -\sqrt{3} \right],\\
F_2 (c_{10},\phi_0-\psi_0)= & \sqrt{2 c_{10}} \left( \sqrt{1+ 2 e^{(\psi_0 - \phi_0)}} -\sqrt{3} \right).
\end{aligned}
\end{equation}
For the 3-ion case with $z_1=1,z_2=1,z_3=-1$, we have
\begin{equation}
\label{Eq66}
\begin{aligned}
F_j (c_{10},c_{20},\phi_0-\psi_0)= & \sqrt{\frac{c_{j0}}{c_{10}+c_{20}}}  \sqrt{2 c_{j0}} \left(e^{\frac{\phi_0 - \psi_0}{2}} -1\right),\\
F_3 (c_{10},c_{20},\phi_0-\psi_0) = & \sqrt{2 (c_{10}+c_{20})} (e^{(\psi_0 - \phi_0)/2} -1),
\end{aligned}
\end{equation}
where $j=1,2$.

Next, at $x=0$, we consider the boundary conditions of the type
\begin{equation}
\label{Eq67}
\begin{aligned}
\psi(0,t) = \psi_0(t), \quad  p_{i}(0,t) = p_{i0}(t), \quad i=1,..,n.
\end{aligned}
\end{equation}
We summarize the results below.\\
\noindent \textit{\textbf{Theorem 2.} Suppose the LEN and NGEN conditions are satisfied, and let $\psi_0(t), {p_{i0}}(t)$ be the given electric potential and ion concentrations at $x=0$ as in (\ref{Eq67}) and let $\psi_1(t), p_{i1}(t)$ be given at $x=1$ for original system (\ref{Eq53}), then for the EN system (\ref{Eq55}) we have the effective boundary conditions  
\begin{equation}
\label{Eq68}
\begin{aligned}
& \ln c_{i0} +z_i \phi_0 + \frac{\epsilon J_{c_i,0} }{D_i} f_i(c_{k0},\phi_0-\psi_0) \\
&= \ln p_{i0} + z_i \psi_0 + o(\epsilon),\\
&\ln c_{i1} +z_i \phi_1 - \frac{\epsilon J_{c_i,1} }{D_i} f_i(c_{k1},\phi_1-\psi_1) \\
&= \ln p_{i1} + z_i \psi_1 + o(\epsilon),
\end{aligned}
\end{equation}
where $i=1,..,n$, $J_{c_i}$ is defined in (\ref{Eq54}), the argument $c_{k0}$ represents a vector $(c_{10},..,c_{n-1,0})$, subscripts 0 and 1 denote quantities at $x=0$ and $x=1$ respectively, and 
\begin{equation}
\label{Eq69}
\begin{aligned}
&f_i(c_{k0},\phi_0-\psi_0) \\
=& \pm \frac{1}{\sqrt{2} c_{i0} } \int_1^{e^{\phi_0 -\psi_0}} \frac{u^{-z_i} -1}{\sqrt{\sum_{k=1}^n c_{k0} (u^{z_k} -1)}} \frac{du}{u}.
\end{aligned}
\end{equation}
Here, the $\pm$ are chosen for the cases $\psi_0\le \phi_0$ and $\psi_0\ge \phi_0$ respectively, but $f_i$ is well-defined around $\phi_0= \psi_0$, and if $f_i$ can be integrated out, the expressions from the two cases are the same. 
}

\noindent \textbf{Remark 8.} For the case $z_1= 1,z_2 =-1$, we will recover the previous formulas in Proposition 2. For the case $z_1=2,z_2=-1$, we get
\begin{equation}
\label{Eq70}
\begin{aligned}
f_1 = &  \frac{\sqrt{2+ e^{\phi_0 - \psi_0}} (1+ 2 e^{\phi_0 - \psi_0} )e^{\frac{3}{2} (\psi_0 - \phi_0)}  - 3\sqrt{3}}{3 \sqrt{2} c_{10}^{3/2}},\\
f_2 = &  \frac{\mathrm{arcsinh} \left(e^{(\phi_0 - \psi_0)/2}/\sqrt{2} \right)  - \mathrm{arccsch}(\sqrt{2})}{ \sqrt{2} c_{10}^{3/2}}.
\end{aligned}
\end{equation}
For the case with $z_1=1,z_2=1,z_3=-1$, we have
\begin{equation}
\label{Eq71}
\begin{aligned}
f_j (c_{10},c_{20},\phi_0-\psi_0)= & \frac{\sqrt{2} (e^{(\psi_0 - \phi_0)/2} -1)}{  c_{j0} \sqrt{c_{10}+c_{20}}}  , \\
f_3 (c_{10},c_{20},\phi_0-\psi_0)= &\frac{\sqrt{2} (e^{(\phi_0 - \psi_0)/2} -1)}{  (c_{10}+c_{20})^{3/2}} ,
\end{aligned}
\end{equation}
where $j=1,2$.

\section{Numerical examples}

\subsection{A steady state problem}

As a first example to verify the preceding effective conditions, we take the steady state problem from Rubinstein \cite[pp. 133-134]{rubinstein1990}, since this problem can be solved analytically with effective conditions. Consider the stationary ionic transport in a unity thick unstirred layer adjacent to an ideally cation-permselective interface, and the PNP system with $\pm1$ ions for $ x \in [0,1]$ is
\begin{equation}
\label{eq54}
\begin{aligned}
-\epsilon^2 \psi'' = p-n, \quad p' + p \psi' = -j, \quad n' -n  \psi' = 0,
\end{aligned}
\end{equation}
together with boundary conditions
\begin{equation}
\label{eq54_1}
\begin{aligned}
& p(0) = n(0) = 1,\quad \psi(0) = 0, \\
& p(1) = 1,\quad J_n(1)=0, \quad \psi(1) = -V,
\end{aligned}
\end{equation}
where prime denotes the derivative with respect to $x$, and $j$ is a flux constant to be determined with given potential $V$. Physically, the $j$-$V$ relation is the current-voltage relation in this example. Since it is electro-neutral at left end $x=0$, there is only a boundary layer near $x=1$.

One can easily write down the EN system for $c(x)$ and $\phi(x)$ in bulk region, and the solutions are given by 
\begin{equation}
\label{eq55}
\begin{aligned}
c(x) = 1-\frac{j}{2} x,\quad \phi(x) = \ln \left( 1-\frac{j}{2} x\right).
\end{aligned}
\end{equation}
By using the usual continuity of electro-chemcial potential as in \cite{rubinstein1990}, we get
\begin{equation}
\label{eq56}
\begin{aligned}
& j = 2\left(1-e^{-V/2}\right).
\end{aligned}
\end{equation}
By the effective condition $(\ref{eq38})$ at $x=1$, we get
\begin{equation}
\label{eq57}
\begin{aligned}
2\ln \left(1-\frac{j}{2}\right) - 4 j \epsilon \left(\frac{\sqrt{2} e^{-\frac{V}{2}}}{(2-j)^2} - \frac{1}{(2-j)^{3/2}}\right) = -V,
\end{aligned}
\end{equation}
where there is an $O(\epsilon)$ correction term. 

\begin{figure}[h]
\begin{center}
\includegraphics[width=3.1in]{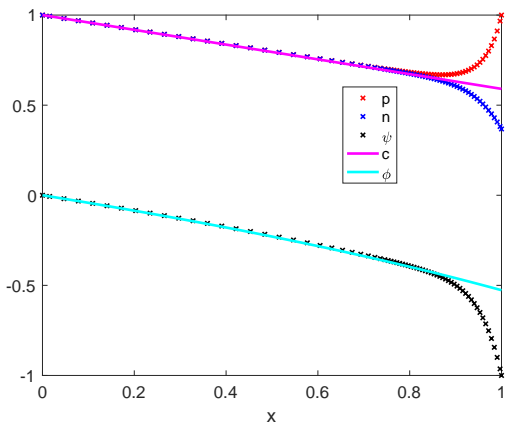}
\caption{\label{fig1} Comparison between analytic bulk solution with numerical solution at $t=20$, with $\epsilon=0.05$.}
\end{center}
\end{figure}

In the numerical verification, we consider the dynamic system (\ref{eq1}) with boundary conditions (\ref{eq54_1}) and the following initial conditions at $t=0$,
\begin{equation}
\label{eq58}
\begin{aligned}
p(x,0 ) = 1,\quad n(x,0)=1.
\end{aligned}
\end{equation} 
The solution tends to the steady state solution of (\ref{eq54}) and (\ref{eq54_1}), and the flux $j$ near the steady state can be found. Finite-volume method with refined mesh near $x=1$ is adopted in the numerical simulation. With $V=1$ and $\epsilon = 0.1, 0.05, 0.01$, we compare in Table \ref{table1} the results of flux $j$ from Rubinstein's condition (\ref{eq56}) and present condition (\ref{eq57}) with that in numerical simulations. The table manifests that the present effective condition produces better results and the $O(\epsilon)$ term is correct. Figure \ref{fig1} shows the comparison between bulk solution (\ref{eq55}) and the numerical solution at $t=20$ with $\epsilon=0.05$.

\begin{table}
\begin{center}
\begin{tabular}{c|c|c|c}
\hline\hline
$\epsilon$ & 0.1 & 0.05& 0.01\\
Rubinstein & 0.7869 \quad& 0.7869 \quad& 0.7869\\
Present & 0.8191 & 0.8029  &  0.7901\\
Numerical & 0.8191 & 0.8029 & 0.7901\\
\hline\hline
\end{tabular}
\caption{\label{table1}Comparison of flux $j$ with fixed $V=1$ and different $\epsilon$, where ``Rubinstein" and ``present" are from formulas  (\ref{eq56}) and  (\ref{eq57}), ``Numerical" is from dynamic system.}
\end{center}
\end{table}

Next, we replace Dirichlet boundary condition for $\psi(1)$ by a Robin-type condition $\eta \psi'(1)= -1-  \psi (1)$ and keep others the same as before. The effective conditions (\ref{eq38},\ref{eq48}) imply that
\begin{equation}
\label{eq59}
\begin{aligned}
&2 \log   \left(1-\frac{j}{2}\right) - \frac{4 j \epsilon  \left( \sqrt{2} e^{\psi_1/2} - \sqrt{2-j}\right)}{(j-2)^2}=\psi_1,\\
&\psi_1 = -1 + \sqrt{2}\frac{\eta}{\epsilon} - 2 W\left(e^{(\sqrt{2}\eta -\epsilon )/(2\epsilon)} \eta/(\sqrt{2}\epsilon)\right),
\end{aligned}
\end{equation} 
where $W$ is the Lambert-W function. For $\epsilon=0.01$ and different $\eta$, we compare in Table \ref{table2} the results of flux $j$ from (\ref{eq59}) with numerical results by a dynamic simulation (with initial condition (\ref{eq58})).  For the Robin-type condition, in general we only get leading order correct, and the error is roughly $O(\eta)$.

\begin{table}
\begin{center}
\begin{tabular}{c|c|c|c}
\hline\hline
$\eta$ & $10^{-2}$ & $10^{-3}$ & $10^{-4}$\\
Present &0.5358\quad & 0.7570 \quad & 0.7867\\
Numerical & 0.5406 & 0.7590 & 0.7871\\
\hline\hline
\end{tabular}
\caption{\label{table2}Comparison of flux $j$ from present formula (\ref{eq59}) and numerical simulation of dynamic system.}
\end{center}
\end{table}

\subsection{Two dynamic examples}

In this subsection, we investigate two dynamical examples to verify the EN theories. In these examples, the previous assumptions on BL structure are satisfied, so we can solve the EN system in bulk region directly and efficiently with effective boundary conditions. 

As the first example, we examine the PNP system (\ref{eq1}) with Dirichlet boundary conditions for $p,n,\psi$ like (\ref{eq25}).  More precisely the boundary conditions at $x=0,1$ are given by
\begin{equation}
\label{eq61}
\begin{aligned}
&p(0,t) =1+t, \quad  n(0,t)= 1, \quad \psi(0,t)= 0, \\
&p(1,t) =1,\quad n(1,t) =1+t,\quad \psi (1,t) = 0,
\end{aligned}
\end{equation}
and the initial conditions at $t=0$ are
\begin{equation}
\label{eq62}
\begin{aligned}
n(x,0) = p(x,0) = 1, \quad 0<x<1.
\end{aligned}
\end{equation} 

In this case, the BL will gradually appear, and we will demonstrate it with $\epsilon = 0.01$. Finite-Volume method with refined mesh in BL is adopted to solve the original system (\ref{eq1}). The EN system (\ref{eq6}) for $c(x,t)$ and $\phi(x,t)$ is solved with fixed uniform mesh with the aid of effective boundary conditions. And we try two implementations with following effective boundary conditions, (i) the leading order Dirichlet conditions (called Rubinstein's condition here)
\begin{equation}
\label{eq63}
\begin{aligned}
&c_0 = c_1 = \sqrt{1+t},\quad \phi_0= -\phi_1 = \frac12 \ln (1+t),
\end{aligned}
\end{equation} 
and (ii) the Robin-type conditions in (\ref{eq40}) with $O(\epsilon)$ correction term.
All the above effective boundaries are explicit and linear and thus can be easily applied. 

\begin{figure}[h]
\begin{center}
\subfigure[ Ion concentrations]{\includegraphics[width=3.1in]{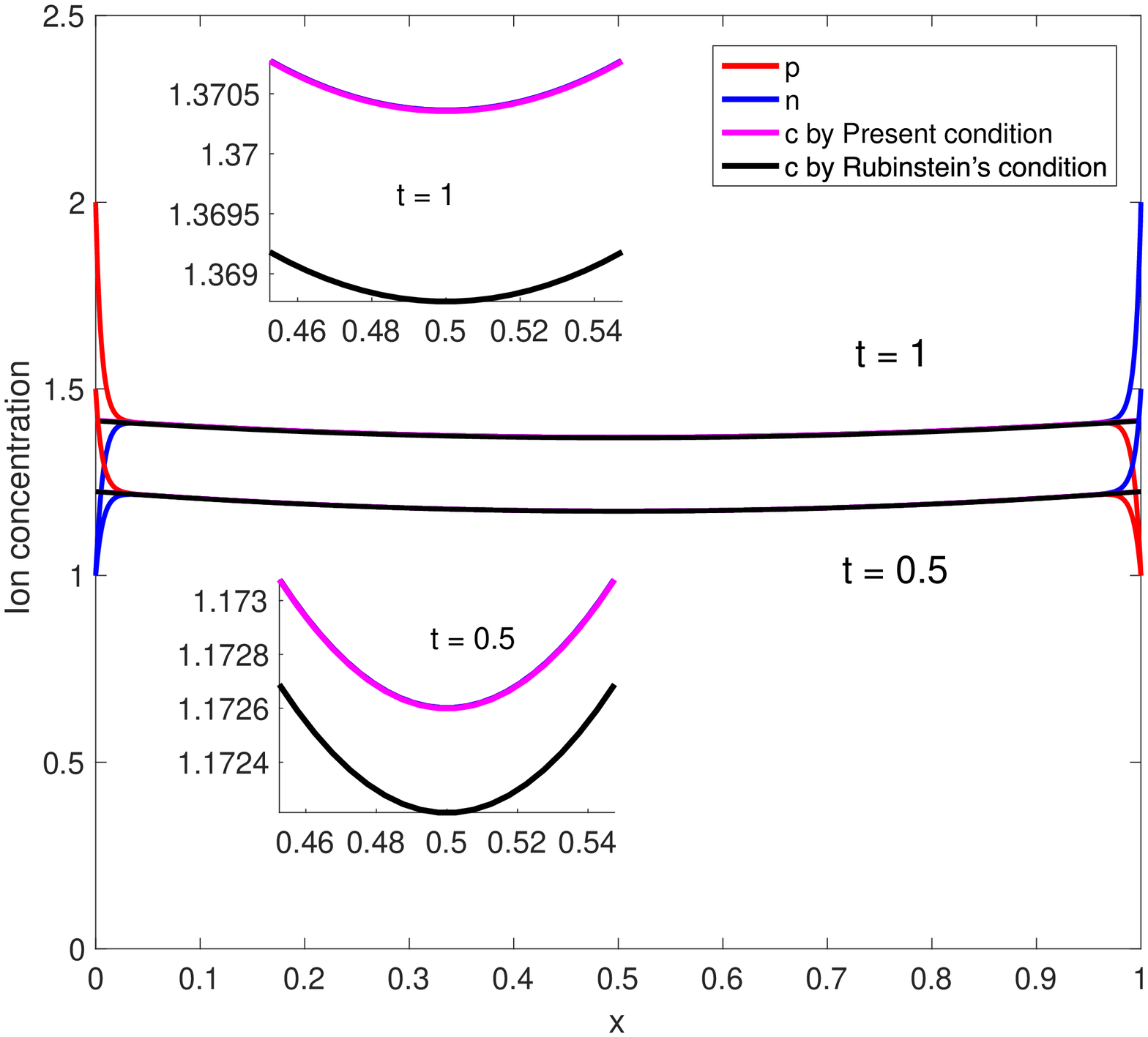}}
\subfigure[ Electric potential]{\includegraphics[width=3.1in]{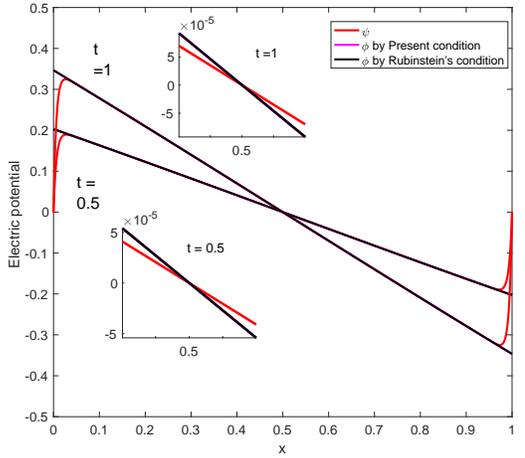}}
\caption{\label{fig2} Comparison between numerical results of original PNP system with Dirichlet conditions and those of the EN system with Rubinstein's condition (\ref{eq63}) and present condition (\ref{eq40}).}
\end{center}
\end{figure}

Figure \ref{fig2} shows the comparison between the numerical result of original PNP system and two direct numerical implementations for EN system. Figure \ref{fig2}(a) shows that the present higher order effective conditions produce better results for ion concentration than leading order Rubinstein's condition, and so does Figure \ref{fig2}(b) for electric potential (note that red and pink curves coincide in the enlarged figure). By using the numerical results of $p(x,t)$ and $\psi(x,t)$ for original system as the exact solution, Table \ref{table3} shows the maximum errors of $c(x,t)$ and $\phi(x,t)$ in some bulk region $[0.25, 0.75]$. This indicates that the accuracy is acceptable with the effective boundary conditions. Furthermore, the EN system allows for relatively large mesh and time step sizes, so the computational time is greatly reduced. For instance, it costs roughly 1 hour to compute the original PNP system up to $t=1$ while it costs 8 minutes for the EN system. 

\begin{table}[h]
\begin{center}
\begin{tabular}{ c|c|c|c|c}
    \hline\hline
 & $c$ at  $t = 0.5$      & $c$ at  $t = 1$        &  $\phi$ at  $t = 0.5$
& $\phi$ at  $t = 1$  \\ \hline
Rubinstein & $4.5 \times10^{-4}$ & $1.7\times 10^{-3}$ & $9.5\times 10^{-5} $
& $6.2\times10^{-5}$\\
Present & $4.9\times 10^{-6}$   &  $8.4\times 10^{-6}$  & $1.1\times 10^{-5}$ 
&  $2.3\times 10^{-5}$\\
$p-n$ & $3.9\times10^{-6}$      & $5.0\times10^{-6}$      & $ - $ & $ - $ \\
\hline\hline
\end{tabular}
\end{center}
\caption {\label{table3} Maximum error in concentration $c(x,t)$ and potential $\phi(x,t)$ in some bulk region $x\in[0.25,0.75]$ and $t=0.5,1$, with Rubinstein's condition (\ref{eq63}) and present condition (\ref{eq40}).}
\end{table}

As a second example, we examine the PNP system (\ref{eq1}) with the flux conditions
\begin{equation}
\label{eq65}
\begin{aligned}
&J_p (0,t)= 0.2, \quad  J_n(0,t) = 0.4, \quad \psi(0,t) = 0, \\
&J_p(1,t) =0.2,\quad J_n(1,t) =0.4(1+ 2 \epsilon ),\quad \psi (1,t) = 0.
\end{aligned}
\end{equation}
Here, the fluxes are $O(1)$ but the unbalanced flux is only $O(\epsilon)$, which is consistent with previous assumptions. The initial conditions are the same as in (\ref{eq62}). The original PNP system for $p,n,\psi$ is simulated by finite volume method with refined mesh, while the EN system (\ref{eq6}) for $c,\phi$ is computed with uniform mesh and the linearized version of the effective boundary conditions (\ref{eq24}). 

\begin{figure}[h]
\begin{center}
\subfigure[ Ion concentrations] {\includegraphics[width=3.1in]{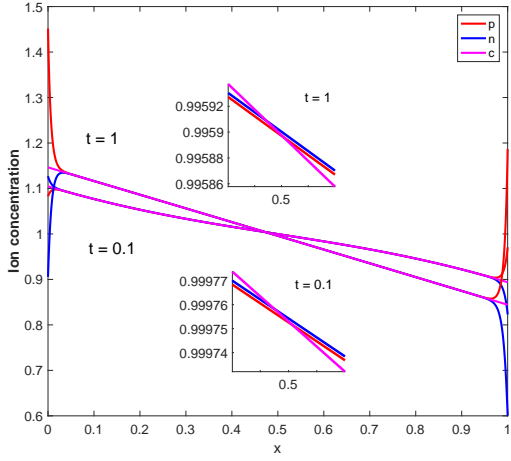}}
\subfigure[ Electric potential]{\includegraphics[width=3.1in]{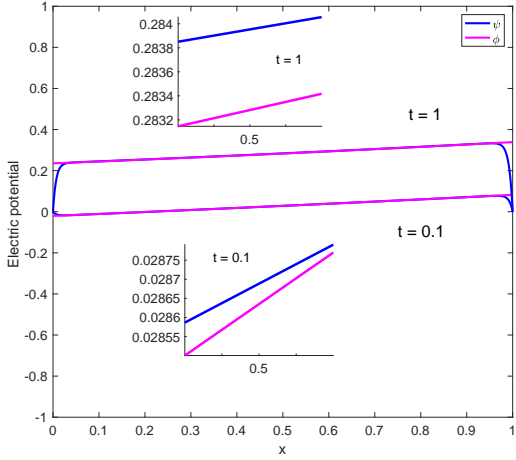}}
\caption{\label{fig3}Comparison between numerical results of original PNP system with flux  conditions and those of the EN system with effective boundary conditions (\ref{eq24}).}
\end{center}
\end{figure}

Figure \ref{fig3} shows the comparison between the numerical results of $p,n,\psi$ from original PNP system and those of $c,\phi$ from EN system at two times $t=0.1$ and 1. It shows that the curves agree very well in the bulk region. By using the numerical results of $p(x,t)$ and $\psi(x,t)$ from original system as the exact solution, Table \ref{table4} shows the maximum errors of $c(x,t)$ and $\phi(x,t)$ in some bulk region $[0.25, 0.75]$.  Similarly, it costs less computational time for EN system with effective boundary conditions, namely 6 minutes compared to about 1 hour for original system.

\begin{table}[h]
\begin{center}
\begin{tabular}{ c|c|c|c|c}
    \hline\hline
 & $c$ at  $t = 0.1$      & $c$ at  $t = 1$        &  $\phi$ at  $t = 0.1$
& $\phi$ at  $t = 1$  \\ \hline
Present & $9.1\times 10^{-7}$   &  $3.6\times 10^{-7}$  & $4.4\times 10^{-5}$ 
&  $6.7\times 10^{-4}$\\
$p-n$ & $2.3\times10^{-6}$      & $3.7\times10^{-6}$      & $ - $ & $ - $ \\
\hline\hline
\end{tabular}
\end{center}
\caption {\label{table4}Maximum error in concentration $c(x,t)$ and potential $\phi(x,t)$ in some bulk region $x\in[0.25,0.75]$ and $t=0.1$ and 1.}
\end{table}

\section{An electro-neutral model for neuronal axon}
\label{sec4}

As a concrete example, we consider a cell structure with a membrane in between \cite{pods2013}. The interval is set to be $[0,L]$ where $L$ is the typical length of cell, and the membrane is in the middle $x=L/2$. The left part $[0,L/2)$ is the extracellular space and  $(L/2,L]$ is the intracellular space. Only three basic ions $\mathrm{Na}^{+},\mathrm{K}^{+},\mathrm{Cl}^{-}$ are considered (fixed negative charge are incorporated into $\mathrm{Cl}^{-}$ ion as approximation), and LEN condition in bulk region is valid in this biological application.

We first formulate the original system in dimensional form. Let $p_i$ ($i=1,2,3$) denote ion concentrations of $\mathrm{Na}^{+},\mathrm{K}^{+},\mathrm{Cl}^{-}$, with valences $z_1=z_2 =1,z_3=-1$. The dimensional PNP system for $p_i$ and electric potential $\psi$ in left region $x\in (0,L/2)$ is 
\begin{equation}
\label{eq83}
\begin{aligned}
&- \epsilon_0 \epsilon_r^L \partial_{xx}\psi  = e_0N_A\left(\sum_{k=1}^3 z_k p_k \right) ,\\
&\partial_t p_i = - \partial_{x} (J_{p_i})= D_i \partial_{x} \left(\partial_{x}p_i +\frac{e_0}{k_B T} z_i p_i \partial_{x}\psi\right),
\end{aligned}
\end{equation}
where $i=1,2,3$, $\epsilon_0$ is vacuum permittivity,  $\epsilon_r^L$ is the relative permittivity of left region (extracellular space), $e_0$  is elementary charge, $N_A$ is Avogadro constant, $D_i$ is the diffusion constant, $k_B$ is Boltzmann constant and $T$ is absolute temperature. The system for right half region is the same except a possibly different relative permittivity ${\epsilon}_r^R$.  The boundary conditions at $x=0,1$ are omitted here, and will be presented in dimensionless form.

The membrane at $x=L/2$ is described by Hodgkin-Huxley (HH) model \cite{hodgkin1990}, in order to simulate action potential for neuronal axon (one might use GHK flux model \cite{hille2001} for other purpuses). Thus, the dimensional relation for the current through membrane/ion channel, from intracellular region to extracellular region, is
\begin{equation}
\label{eq84}
\begin{aligned}
& I_i =  G_{p_i} (V_m - E_i),
\end{aligned}
\end{equation}
or in terms of flux at $x=L/2$
\begin{equation}
\label{eq85}
\begin{aligned}
-z_i e_0 N_A J_{p_i} =&z_i e_0 N_A D_i  \left(\partial_{x}p_i +\frac{e_0}{k_B T} z_i p_i \partial_{x}\psi\right) \\
=& G_{p_i} \left(\psi_R - \psi_L -  \frac{k_B T}{z_i e_0} \ln \frac{p_{iL}}{p_{iR}}\right),
\end{aligned}
\end{equation}
where $G_{p_i}$ is the conductance for ion $p_i$, $V_m = \psi_R-\psi_L$ is the membrane potential, $E_i$ is the Nernst potential of ion $p_i$, subscripts $L$ and $R$ denote the left and right limits of quantities at membrane $x={L}/{2}$.

Suppose the membrane has thickness $h_m$ and relative permittivity $\epsilon_r^m$, and assume there are no ions in membrane. Thus, the electric potential is linear inside membrane. The other two jump conditions on the interface $x=L/2$ are
\begin{equation}
\label{eq86}
\begin{aligned}
& \left.\epsilon_r^L \partial_{x}\psi \right|_{x=\frac{L}{2}-}  =\left.\epsilon_r^R \partial_{x}\psi \right|_{x=\frac{L}{2}+} = \epsilon_r^m \frac{\psi_R - \psi_L}{h_m},
\end{aligned}
\end{equation}
where $\frac{L}{2}\pm$ mean the left and right limits at $x=L/2$.

The conductances depend on membrane potential $V_m$. Following \cite{pods2013}, we set $G_{p_3}\equiv G_{{Cl}} =0$ and
\begin{equation}
\label{eq87}
\begin{aligned}
& G_{p_1} \equiv G_{{Na}} = \bar{G}_{Na} m^3 h +G_{Na,leak}, \\
& G_{p_2} \equiv G_{{K}} = \bar{G}_{K} n^4 + G_{K,leak},
\end{aligned}
\end{equation}
where
\begin{equation}
\label{eq88}
\begin{aligned}
& \frac{dn}{dt} = \alpha_n (1-n) -\beta_n n,\\
& \frac{dm}{dt} = \alpha_m (1-m) -\beta_m m,\\
& \frac{dh}{dt} = \alpha_h (1-h) -\beta_h h.\\
\end{aligned}
\end{equation}
The coefficients depend on $V_m$ and are given by
\begin{equation}
\label{eq89}
\begin{aligned}
& \alpha_n =\frac{1}{100} \frac{10-\bar{V}}{\left(e^{(10-\bar{V})/10} -1\right)},\quad 
\beta_n = \frac{1}{8 e^{\bar{V}/80}},\\
& \alpha_m = \frac{1}{10} \frac{25-\bar{V}}{\left(e^{(25-\bar{V})/10} -1\right)}, \quad \beta_m = 4 e^{-\bar{V}/18},\\
& \alpha_h= \frac{7}{100} e^{-\bar{V} /20}, \quad \beta_h = \frac{1}{e^{(30-\bar{V})/10}+1},
\end{aligned}
\end{equation}
where $\bar{V} = V_m - V_r$ and $V_r$ is some fixed resting potential. In above coefficients, the unit for $\bar{V}$ is millivolt. Theoretically, there is no singularity in above coefficients, but for computation when $\bar{V}$ is near $10$ or $25$, it is sensitive as denominator approaches 0. We can use the Taylor expansions in a small neighbourhood say $\delta=0.01$,
\begin{equation}
\label{eq90}
\begin{aligned}
& \alpha_n(\bar{V} ) = \frac{1}{10} + \frac{\bar{V} -10}{200}+ \frac{(\bar{V} -10)^2}{12000},\quad |\bar{V}-10|<\delta,\\
& \alpha_m(\bar{V} ) =1+ \frac{\bar{V} -25}{20} + \frac{(\bar{V} -25)^2}{1200},\quad |\bar{V}-25|<\delta,
\end{aligned}
\end{equation}
and the error by choosing $\delta=0.01$ is at least at the order of $10^{-12}$.
With $\bar{V}=0$, we get the steady state (also used as initial values to simulate action potential)
\begin{equation}
\label{eq91}
\begin{aligned}
& n_0 = \frac{4}{5e -1} \approx 0.3177,\quad m_0 = \frac{5}{8 e^{5/2} -3} \approx 0.05293,\\
& h_0= \frac{7 (1+e^3)}{107 + 7 e^3} \approx 0.5961.
\end{aligned}
\end{equation}

\subsection{Non-dimensionalization}

In this subsection, we present the dimensionless PNP formulation combined with HH model.
We adopt the following scalings 
\begin{equation}
\label{eq92}
\begin{aligned}
& \tilde{\psi} = \frac{\psi}{k_B T/e_0}, \quad \tilde{p}_i = \frac{p_i}{p_0},\quad \tilde{x} = \frac{x}{L}, \quad \tilde{h}_m = \frac{h_m}{L}, \\
& \tilde{D}_i = \frac{D_i}{D_0}, \quad \tilde{t} = \frac{t}{L^2/D_0},\quad \tilde{G}_{p_i}= \frac{G_{p_i}}{G_0},
\end{aligned}
\end{equation}
where $p_0$ is the typical concentration of ions, $D_0$ is the typical diffusion constant, and typical conductance $G_0$ is defined by $G_0 = p_0 D_0 e^2 N_A/(k_B T L)$. All the parameter values and typical values are given in Appendix \ref{appendix_B}. In the following, we will remove the tilde, and still use the same notations but they represent dimensionless quantities.

The dimensionless PNP system is given by
\begin{equation}
\label{eq93}
\begin{aligned}
&- \epsilon_L^2 \partial_{xx}\psi  = \sum_{k=1}^3 z_k p_k, \quad 0<x<1/2,\\
&- \epsilon_R^2 \partial_{xx}\psi  = \sum_{k=1}^3 z_k p_k ,\quad 1/2<x<1,\\
&\partial_t p_i = - \partial_{x} (J_{p_i})=  D_i \partial_{x} \left(\partial_{x} p_i + z_i p_i \partial_{x} \psi \right), \\
&\qquad \qquad i=1,2,3 \quad  0<x<1,
\end{aligned}
\end{equation}
together with the conditions on interface $x=1/2$, 
\begin{equation}
\label{eq94}
\begin{aligned}
\left.-z_i J_{p_i} \right|_{x=\frac12} \equiv  &\left.z_i D_i  \left(\partial_{x}p_i + z_i p_i \partial_{x}\psi\right)  \right|_{x=\frac12}\\
=& G_{p_i} \left(\psi_R - \psi_L -  \frac{1}{z_i} \ln \frac{p_{iL}}{p_{iR}}\right),
\end{aligned}
\end{equation}
and
\begin{equation}
\label{eq95}
\begin{aligned}
& \left.\epsilon_L^2\partial_{x}\psi \right|_{x=\frac{1}{2}-}  =\left.\epsilon_R^2 \partial_{x}\psi \right|_{x=\frac{1}{2}+} = \epsilon_m^2 \frac{\psi_R -\psi_L}{h_m}.
\end{aligned}
\end{equation}
In the above, the dimensionless parameters $\epsilon_L, \epsilon_R, \epsilon_m$ are defined by
\begin{equation}
\label{eq96}
\begin{aligned}
& \epsilon_s \equiv \sqrt{\frac{\epsilon_0 \epsilon_r^s k_B T}{e^2 N_A p_0 L^2}}, \quad s= L,R,m.
\end{aligned}
\end{equation}
The values of above dimensionless parameters are given in Appendix \ref{appendix_B}.

We use some typical bulk concentrations as the initial state (see Appendix \ref{appendix_B}), so we have at $t=0$,
\begin{equation}
\label{eq97}
\begin{aligned}
& p_1(x,0) = 1,\quad p_2(x,0) = 0.04,\quad p_3 (x,0) = 1.04, \\
& \qquad \mathrm{for} \quad 0<x<1/2,\\
& p_1(x,0) = 0.12,~~ p_2(x,0) = 1.25,~~p_3 (x,0) = 1.37, \\
& \qquad \mathrm{for} \quad1/2<x<1.\\
\end{aligned}
\end{equation}
Regarding boundary conditions at $x=0,1$, we adopt
\begin{equation}
\label{eq98}
\begin{aligned}
&\psi (0,t) = 0,\quad p_1(0,t) = 1,\\
& p_2(0,t) = 0.04,\quad p_3 (0,t) = 1.04,\\
&\frac{\partial \psi}{\partial x}(1,t) = 0,\quad  J_{p_i}(1,t) =0, \quad i=1,2,3. 
\end{aligned}
\end{equation}
where the first two lines mean fixed concentrations and electric potential in extracellular region, and the third line means the symmetric conditions in the middle of neuronal axon.

This system is calculated together with definition (\ref{eq87}) for conductances and the dynamics of $n,m,h$ in (\ref{eq88}). We will not scale the quantities in the coefficients (\ref{eq89}), where the quantity $\bar{V}$ (in millivolts) is related to normalized membrane potential $V_m=\psi_R-\psi_L$ through
\begin{equation}
\label{eq99}
\begin{aligned}
& \bar{V} =  \frac{k_B T}{e_0}(\psi_R-\psi_L) - V_r,
\end{aligned}
\end{equation}
where $V_r$ is fixed resting potential in millivolts (see (\ref{eq107}) below).

\subsection{The effective flux conditions for the electro-neutral model}

The reduced EN model for bulk region consists of the equations in (\ref{Eq55}) for $c_1,c_2, \phi$, with $n=3$ and $z_1=z_2=1,z_3=-1$. Also, we need to propose approximate jump conditions at middle interface for bulk quantities $c_{iL},\phi_L,c_{iR},\phi_R$ ($i=1,2$), where subscripts $L,R$ indicate the left and right limits of quantities at interface $x=1/2$.  Based on previous analysis in Subsections \ref{sec2_4} and \ref{sec2_5}, we derive the following 14 conditions to close the system (normally 6 conditions are needed, but we have introduced 8 auxiliary quantities $p_{iL},p_{iR},\psi_L,\phi_R$ with $i=1,2,3$ at interface $x=1/2$)
\begin{equation}
\label{eq100}
\begin{aligned}
&\frac{\epsilon_m^2}{h_m}({\psi_R - \psi_L}) \\
=&  \epsilon_R \sqrt{2 c_{3R}} \left(e^{(\phi_R - \psi_R)/2} -  e^{(\psi_R - \phi_R)/2}\right)\\
=& -\epsilon_L\sqrt{2 c_{3L}} \left(e^{(\phi_L - \psi_L)/2} -  e^{(\psi_L - \phi_L)/2}\right),
\end{aligned}
\end{equation}
where $c_{3R} = c_{1R} + c_{2R},c_{3L} = c_{1L} + c_{2L}$ by electro-neutrality condition, and
\begin{equation}
\label{eq101}
\begin{aligned}
&G_{p_i} \left(\psi_R - \psi_L -  \frac{1}{z_i} \ln \frac{p_{iL}}{p_{iR}}\right) \\
 = &- z_i \left( J_{c_i,R} + \epsilon_R \partial_t F_{iR}\right)\\
=& - z_i \left( J_{c_i,L} - \epsilon_L \partial_t F_{iL}\right),\\
&\ln c_{iR} + z_i \phi_R + \frac{\epsilon_R J_{c_i,R} }{D_i} f_{iR}
= \ln p_{iR} + z_i \psi_R , \\
&\ln c_{iL} + z_i \phi_L - \frac{\epsilon_L J_{c_i,L} }{D_i} f_{iL} = \ln p_{iL} + z_i \psi_L,
\end{aligned}
\end{equation}
where $i=1,2,3$, $J_{c_i}$ is given in (\ref{Eq54}), and we have defined
\begin{equation}
\label{eq102}
\begin{aligned}
 & F_{is}= F_i(c_{1s},c_{2s},\phi_s-\psi_s),  \\
 & f_{is}= f_i(c_{1s},c_{2s},\phi_s-\psi_s), \quad s=L,R,
\end{aligned}
\end{equation}
where $F_i$ and $f_i$ are given by (\ref{Eq66}) and (\ref{Eq71}).

From definition (\ref{eq87}) and the data in Appendix \ref{appendix_B}, the conductances $G_{p_i}$ have at most the same order as the dimensionless small parameter $\epsilon_R=\epsilon_L$. Now we simplify the conditions in (\ref{eq101}) asymptotically and get 
\begin{equation}
\label{eq103}
\begin{aligned}
& - z_i J_{c_i,R}   \\
=&  G_{p_i} \left(\phi_R - \phi_L -  \frac{1}{z_i} \ln \frac{c_{iL}}{c_{iR}}\right) + z_i \epsilon_R \partial_t F_{iR} \\
 & + \frac{G_{p_i} \epsilon_R}{D_i z_i} \left[ J_{c_i,R} f_{iR} + \frac{\epsilon_R}{\epsilon_L}J_{c_i,L} f_{iR}  \right],\\
=&  G_{p_i} \left(\phi_R - \phi_L -  \frac{1}{z_i} \ln \frac{c_{iL}}{c_{iR}}\right) \\
&+ z_i \epsilon_R \partial_t F_{iR} +O(\epsilon_R^2).
\end{aligned}
\end{equation}
We expect the flux is $O(\epsilon_R)$, so the higher-order term $O( \epsilon_R^2)$ can be ignored in calculation but the term $\epsilon_R \partial_t F_{iR}$ should be kept. Similarly, we have the other condition for bulk flux $J_{c_i,L}$ at interface 
\begin{equation}
\label{eq104}
\begin{aligned}
 - z_i J_{c_i,L}  =&  G_{p_i} \left(\phi_R - \phi_L -  \frac{1}{z_i} \ln \frac{c_{iL}}{c_{iR}}\right) \\
 &- z_i \epsilon_L \partial_t F_{iL}+ +O(\epsilon_L^2).
\end{aligned}
\end{equation}
Formally, without the $F_{iR},F_{iL}$ corrections, these two conditions are exactly the same as HH model for bulk quantities in EN model, but $F_{iR},F_{iL}$ are in general not negligible.

To summarize, the final EN system consists of equations (\ref{Eq55}), jump conditions (\ref{eq100},\ref{eq103},\ref{eq104}), and dynamics of conductances (\ref{eq87}-\ref{eq89}).  

\noindent\textbf{Remark 9.} If we carry out a linearization for small $\phi_L - \psi_L$ and $\phi_R - \psi_R$ in exponential functions (in $F_i,f_i$ and (\ref{eq100})), then we obtain
\begin{equation}
\label{eq105}
\begin{aligned}
 & -\sum_{i=1}^3 z_i J_{c_i,L} - \sum_{i=1}^3 G_{p_i} \left(\phi_R - \phi_L -  \frac{1}{z_i} \ln \frac{c_{iL}}{c_{iR}}\right)\\
 & = - \epsilon_L \sum_{i=1}^3  z_i \partial_t F_{iL} \approx  -\epsilon_L \partial_t [\sqrt{2 c_{3L}}(\phi_L - \psi_L )] \\
  & \approx \frac{\epsilon^2_m}{h_m } \partial_t(\psi_R- \psi_L),
\end{aligned}
\end{equation}
where we have used linearized version of (\ref{eq100}) in last approximation. Similarly, we get
\begin{equation}
\label{eq106}
\begin{aligned}
 & -\sum_{i=1}^3 z_i J_{c_i,R} - \sum_{i=1}^3 G_{p_i} \left(\phi_R - \phi_L -  \frac{1}{z_i} \ln \frac{c_{iL}}{c_{iR}}\right)  \\
 &\approx \frac{\epsilon^2_m}{h_m } \partial_t(\psi_R- \psi_L).
\end{aligned}
\end{equation}
Physically, on the left-hand side, the first term is the total current from/to bulk region, where the minus sign means from intracellular space to extracellular space; and the second term is total current through membrane or ion channels, approximated by bulk quantities (they differ by higher order term). On the right-hand side, the quantity ${\epsilon^2_m}/{h_m }$ is the normalized membrane capacitance (scaled by ${e^2 N_Ap_0 L}/{(k_B T)}$). Under such linearization, this equation reduces to the HH cable model. It is worth noting the electro-neutral model \cite{mori2009} proposed by Yoichiro Mori in a 3D framework. The present formulation shares similar form with his model, Eq.32-Eq.35 therein. He introduced the term $\sigma_i$ in Eq.35 which serves the same role as $F_{iR},F_{iL}$ above.

\subsection{Numerical results}

In this subsection, we show some numerical results for both original model with PNP system and the present EN model. The simulation is divided into two steps, first we generate a resting state and second we simulate the phenomenon of action potential.

\begin{figure}[h]
\begin{center}
\subfigure[The dynamics of membrane potential $V_m$] {\includegraphics[width=2.6in]{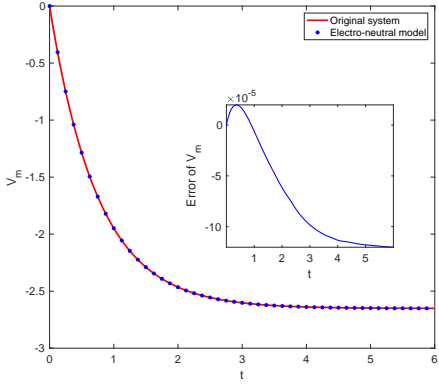}}
\subfigure[Distribution of electric potential at resting state]{\includegraphics[width=2.6in]{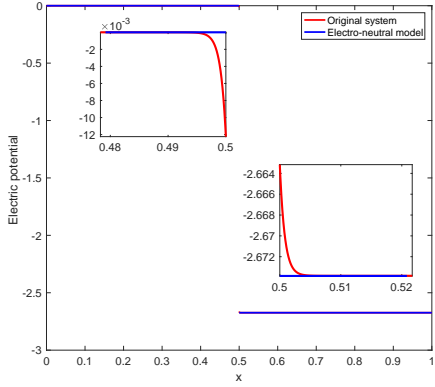}}
\caption{\label{fig4} Numerical results of original system and electro-neutral (EN) model to generate the resting state in step 1.}
\end{center}
\end{figure}

Step 1. To generate a resting state, we only use two leak conductance as in \cite{pods2013}, by setting $\bar{G}_{Na} =\bar{G}_{K}=0$ in (\ref{eq87}). Flux of sodium ion is positive, i.e., from extracellular region to intracellular region, while flux of potassium ion is negative. After some time, say at $t=6$, the net flux across membrane tends to 0, i.e., $\left.J_{p_1}+ J_{p_2}\right|_{x=1/2} =0$, which is defined as the resting state. Figure \ref{fig4}(a) shows the dynamics of membrane potential $V_m=\psi_R-\psi_L$ for both original model and EN model, and they agree very well. Figure \ref{fig4}(b) shows the distributions of electric potential $\psi$ for original system and $\phi$ for EN model, at resting state $t=6$. They agree very well in bulk region. The resting potential is calculated as 
\begin{equation}
\label{eq107}
\begin{aligned}
& \left.V_m\right|_{t=6} =\left.\psi_R-\psi_L\right|_{t=6} \approx -2.65,\\
& V_r = \frac{k_B T}{e_0} \left.(\psi_R-\psi_L)\right|_{t=6}  \approx  -63.8 \, \mathrm{mV}.
\end{aligned}
\end{equation}
The EN model needs less mesh points and allows a relatively large time step, therefore it is more efficient in computation. To calculate the resting state by original system, it costs roughly 3.7 hours, while it takes only 4 minutes by the EN model. 

\begin{figure}[h]
\begin{center}
\subfigure[With $\Delta t= 5*10^{-6}$]{\includegraphics[width=2.6in]{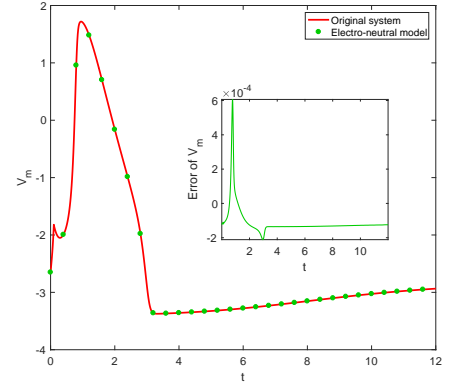}}
\subfigure[With $\Delta t= 5*10^{-5}$]{\includegraphics[width=2.6in]{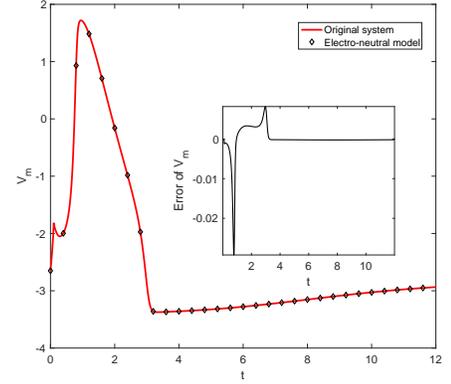}}
\subfigure[With $\Delta t= 5*10^{-4}$] {\includegraphics[width=2.6in]{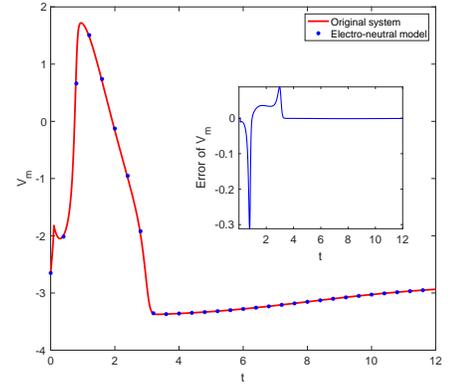}}
\caption{\label{fig5} Numerical results of the original PNP system and EN model in the dynamics of action potential in step 2.}
\end{center}
\end{figure}

Step 2. To evoke the action potential \cite{pods2013}, we use both leak conductance and voltage-gated conductances, with initial values in (\ref{eq91}) for the dynamics of $n,m,h$. To speed the process, we modify $\bar{G}_{Na}$ to allow more flux of $\mathrm{Na}^+$ into the intracellular region. In the simulation, we change $\bar{G}_{Na}$ to $50 \bar{G}_{Na}$ for the period $0<t< 0.1$. Figure \ref{fig5} shows dynamics of membrane potential $V_m=\psi_R-\psi_L$ for both original model and EN model with different time steps, which demonstrates the phenomenon of action potential. Table \ref{table5} compares the computational time between the original system and EN model, and shows the maximum error for the whole process of action potential with different time steps (numerical results from original system is treated as exact value). This indicates that the EN model is more efficient with acceptable error.

%\begin{table*}
%\caption{\label{tab:table3}This is a wide table that spans the full page
%width in a two-column layout. It is formatted using the
%\texttt{table*} environment. It also demonstates the use of
%\textbackslash\texttt{multicolumn} in rows with entries that span
%more than one column.}
%\begin{ruledtabular}
%\begin{tabular}{ccccc}
% &\multicolumn{2}{c}{$D_{4h}^1$}&\multicolumn{2}{c}{$D_{4h}^5$}\\
% Ion&1st alternative&2nd alternative&lst alternative
%&2nd alternative\\ \hline
% K&$(2e)+(2f)$&$(4i)$ &$(2c)+(2d)$&$(4f)$ \\
% Mn&$(2g)$\footnote{The $z$ parameter of these positions is $z\sim\frac{1}{4}$.}
% &$(a)+(b)+(c)+(d)$&$(4e)$&$(2a)+(2b)$\\
% Cl&$(a)+(b)+(c)+(d)$&$(2g)$\footnotemark[1]
% &$(4e)^{\text{a}}$\\
% He&$(8r)^{\text{a}}$&$(4j)^{\text{a}}$&$(4g)^{\text{a}}$\\
% Ag& &$(4k)^{\text{a}}$& &$(4h)^{\text{a}}$\\
%\end{tabular}
%\end{ruledtabular}
%\end{table*}

\noindent \textbf{Remark 10.} In this case, across the membrane, we have Robin-type boundary conditions for $\psi$ in (\ref{eq95}) with an effective $\eta$ as in Subsection \ref{sec2_4}
\begin{equation}
\begin{aligned}
\eta = \frac{\epsilon_L^2 h_m}{\epsilon_m^2} \approx 0.2 \gg O(\epsilon).
\end{aligned}
\end{equation}
As expected in previous Remark 6 (this 3-ion case is similar to 2-ion case with $\pm1$ valences), the variation of $\psi,p_i$ is roughly $O(\epsilon_L/\eta) \sim 10^{-2}$, which is consistent with numerical simulations like Figure \ref{fig4}(b).

\begin{table*}
\begin{center}
\begin{tabular}{ c|c|c|c|c}
    \hline\hline
 & Original system $\Delta t=\frac{5}{8}*10^{-6}$  & EN model, $\Delta t=5*10^{-6}$  &  EN model, $\Delta t=5*10^{-5}$ 
& EN model, $\Delta t=5*10^{-4}$   \\ \hline
Error & $ - $   &  $6*10^{-4}$  & $0.03$ 
&  $0.3$\\
Time & 8 hours      & 1.7 hour      &  10 minutes & 1 minute \\
\hline
\end{tabular}
\end{center}
\caption {\label{table5}Comparison of computation time between original system and EN model, and the maximum error for membrane potential $V_m$ in EN model.}
\end{table*}

\section{Conclusions}

In this work, we have investigated the 1D dynamic PNP system with various boundary conditions, and derived the corresponding EN system with effective boundary conditions. For the case of Dirichlet boundary conditions, the effective conditions can be considered as generalization of continuity of electrochemical potential. For flux conditions, we derived a physically correct effective conditions by bringing back an essential higher-order contribution. The effective conditions for the general multi-ion species case involves elliptic integrals, and these extra dynamic terms of elliptic integrals account for the accumulation of ions in BL. We have validated our EN models with numerical examples and demonstrated the effectiveness of the EN system with the implementation of the well-known Hodgkin-Huxley model for generating action potential on a cell membrane. 

As a next step, we will extend our approach to 2D in the near future, and illustrate the signal transmission in neuronal axon in a more realistic framework. We also plan to extend our approach to modified PNP system where size effect of the ions are included.

\begin{acknowledgments}
The authors would like to thank Prof. Yoichiro Mori for valuable suggestions. Part of the research is supported by NSERC and the Fields Institute.% put your acknowledgments here.
\end{acknowledgments}

\bibliography{reference_pnp}

\appendix

\section{The functions $F_i,f_i$ in Theorems 1,2}
\label{appendix_A}

We show that the functions $F_i,f_i$ are well defined. We take $F_i$ for example, and recall that
\begin{equation}
\begin{aligned}
F_i= \pm \frac{c_{i0}}{\sqrt{2}} \int_1^{e^{\phi_0 -\psi_0}} \frac{u^{z_i} -1}{\sqrt{\sum_{k=1}^n c_{k0} (u^{z_k} -1)}} \frac{du}{u}.
\end{aligned}
\end{equation}
We analyze the integrand, and since all $z_k$ are integers, we can write the numerator as
\begin{equation}
\begin{aligned}
u^{z_i} -1= \frac{P(u)}{Q(u)} (u-1),
\end{aligned}
\end{equation}
where $u=1$ is a simple zero, and polynomials ${P(u)}$ and ${Q(u)}$ are well defined near $u=1$. We write the function inside square root as
\begin{equation}
\begin{aligned}
H(u) = \sum_{k=1}^n  c_{k0} (u^{z_k} -1),
\end{aligned}
\end{equation}
then, one easily finds (note $z_k\ne 0$)
\begin{equation}
\begin{aligned}
& H(1) =0, \quad H'(u) = \sum_{k=1}^n  c_{k0} z_k u^{z_k-1} ,\\
& H'(1)  = \sum_{k=1}^n  c_{k0} z_k=0, \\
& H''(1) = \sum_{k=1}^n  c_{k0} z_k(z_k-1) =\sum_{k=1}^n  c_{k0} z_k^2 >0.
\end{aligned}
\end{equation}
Therefore, $u=1$ is a double zero of $H(u)$. Since all $z_k$ are integers, $H(u)$ is a rational function in $u$, and we can write
\begin{equation}
\begin{aligned}
& H(u) = \frac{P_1(u)}{Q_1(u) } (u-1)^2,
\end{aligned}
\end{equation}
where polynomials ${P_1(u)}$ and ${Q_1(u) }$ are well-defined and ${P_1(u)}/{Q_1(u) }>0$ near $u=1$. Then the integrand together with $\pm$ is given by
\begin{equation}
\begin{aligned}
&\pm \frac{u^{z_i} -1}{u \sqrt{\sum_{k=1}^n c_{k0} (u^{z_k} -1)}} \\
=& \frac{ \pm(u-1)}{\sqrt{(u-1)^2}} \frac{P(u)}{u Q(u)} \sqrt{\frac{Q_1(u)}{P_1(u)}} =\frac{P(u)}{u Q(u)} \sqrt{\frac{Q_1(u)}{P_1(u)}},
\end{aligned}
\end{equation}
where $\pm $ are chosen for cases $u>1$ and $u<1$, and hence the first factor disappears. This implies that $F_i$ has the same form for both cases.

\section{The data used in Section \ref{sec4}}
\label{appendix_B}

The data are mainly from papers \cite{hodgkin1990,pods2013} and the book \cite{malmivuo1995}. The temperature in \cite{hodgkin1990} is set to be $6.3$ $^o C$, so  we get $T = 279.45 \, \mathrm{K}$. The other constants are
\begin{equation}
\begin{aligned}
& k_B  = 1.38 \times 10^{-23} \,\mathrm{J}/\mathrm{K}, \quad N_A = 6.022 \times 10^{23} /\mathrm{mol}, \\
& e_0= 1.602 \times 10^{-19} \,\mathrm{C},~ \epsilon_0 = 8.854 \times 10^{-12} \, \mathrm{C}/(\mathrm{V\cdot m}).
\end{aligned}
\end{equation}
The typical bulk concentrations for $\mathrm{Na}^+,\mathrm{K}^+, \mathrm{Cl}^-$ are
\begin{center}
\begin{tabular}{cccc}
 & $p_1,\mathrm{Na}^+$  & \quad $p_2,\mathrm{K}^+$ & \quad $p_3,\mathrm{Cl}^-$\\
Extracellular & $100\, \mathrm{mM}$ & $4\, \mathrm{mM}$ &$104\, \mathrm{mM}$\\
Intracellular & $12\, \mathrm{mM}$ & $125\, \mathrm{mM}$  &  $137\, \mathrm{mM}$\\
\end{tabular}
\end{center}
\noindent which are used as initial conditions (scaled by $p_0$ below). Some typical values are (diffusivity of $\mathrm{Cl}^-$ is from \cite{bob2014})
\begin{equation}
\begin{aligned}
& \epsilon_r^L = \epsilon_r^R =80, \quad \epsilon_r^m= 2,\quad h_m = 5 \mathrm{nm},\\
& p_0 = 100\, \mathrm{mM}= 100 \, \mathrm{mol}/\mathrm{m}^3, \quad L = 1 \mu \mathrm{m},\\
& D_0 = 10^{-5} \, \mathrm{cm}^2/\mathrm{s}= 10^{-9} \, \mathrm{m}^2/\mathrm{s}, \\
&D_1 = 1.33 D_0,\quad D_2 = 1.96 D_0,\quad D_3 = 2.03 D_0.
\end{aligned}
\end{equation}
The conductances are given by
\begin{equation}
\begin{aligned}
& \bar{G}_{Na} = 120\, \mathrm{mS}/\mathrm{cm}^2 = 1200 \, \mathrm{C}/(\mathrm{V\cdot s \cdot m^2}),\\
& \bar{G}_{K} = 360 \, \mathrm{C}/(\mathrm{V\cdot s \cdot m^2}),\\
& \bar{G}_{Na,leak} = 0.65 \, \mathrm{C}/(\mathrm{V\cdot s \cdot m^2}),\\
& \bar{G}_{K,leak} = 4.35 \, \mathrm{C}/(\mathrm{V\cdot s \cdot m^2}).
\end{aligned}
\end{equation}
From the above data, we get
\begin{equation}
\begin{aligned}
&\frac{k_B T}{e_0} \approx 24  \, \mathrm{mV} ,\quad \frac{L^2}{D_0} = 1\, \mathrm{ms},\\
& G_0= \frac{p_0 D_0 e^2 N_A}{k_B T L} \approx 400758 \, \mathrm{C}/(\mathrm{V\cdot s \cdot m^2}).
\end{aligned}
\end{equation}
For the dimensionless system we have 
\begin{equation}
\begin{aligned}
&\epsilon_L= \epsilon_R= 1.33\times 10^{-3},\quad \epsilon_m =2.1\times 10^{-4}, \\
& h_m = 5 \times 10^{-3},\\
& D_1 = 1.33,\quad D_2 = 1.96, \quad D_3= 2.03,\\
& \bar{G}_{Na} = 3  \times 10^{-3},\quad  \bar{G}_{K} =9  \times 10^{-4},\\
& \bar{G}_{Na,leak} = 1.6 \times 10^{-6},\quad \bar{G}_{K,leak} = 1 \times 10^{-5}.
\end{aligned}
\end{equation}

\end{document}